\documentclass[%
aip,
amsmath,amssymb,
reprint,%
]{revtex4-1}

\usepackage{graphicx} % Required for inserting images
\usepackage{dcolumn}% Align table columns on decimal point
\usepackage{setspace}
\usepackage{bm}% bold math

\usepackage[utf8]{inputenc}
\usepackage[T1]{fontenc}
\usepackage{mathptmx}
\usepackage{etoolbox}

\usepackage{textcomp, gensymb}
\usepackage{orcidlink}
\usepackage{hyperref}
\hypersetup{colorlinks,citecolor=black,filecolor=red,linkcolor=black,urlcolor=blue} 

\newcommand{\mt}{\textcolor{black}}
\newcommand{\OH}{CaOH$^+$}

\makeatletter
\def\@email#1#2{%
 \endgroup
 \patchcmd{\titleblock@produce}
  {\frontmatter@RRAPformat}
  {\frontmatter@RRAPformat{\produce@RRAP{*#1\href{mailto:#2}{#2}}}\frontmatter@RRAPformat}
  {}{}
}%
\makeatother

\begin{document}

\title{Photodissociation spectra of single trapped CaOH$^+$ molecular ions}

\author{Zhenlin Wu
\orcidlink{0000-0002-8188-7701}}
\affiliation{Institut f\"ur Experimentalphysik, Universit\"at Innsbruck, Technikerstraße 25/4, 6020 Innsbruck, Austria}

\author{Stefan Walser \orcidlink{0009-0001-0217-5117}}
\affiliation{Institut f\"ur Experimentalphysik, Universit\"at Innsbruck, Technikerstraße 25/4, 6020 Innsbruck, Austria}

\author{Verena Podlesnic}
\affiliation{Institut f\"ur Experimentalphysik, Universit\"at Innsbruck, Technikerstraße 25/4, 6020 Innsbruck, Austria}

\author{Mariano Isaza-Monsalve
\orcidlink{0009-0006-6996-7916}}
\affiliation{Institut f\"ur Experimentalphysik, Universit\"at Innsbruck, Technikerstraße 25/4, 6020 Innsbruck, Austria}

\author{Elyas Mattivi
\orcidlink{0009-0008-9681-6667}}
\affiliation{Institut f\"ur Experimentalphysik, Universit\"at Innsbruck, Technikerstraße 25/4, 6020 Innsbruck, Austria}

\author{Guanqun Mu}
\affiliation{Institut f\"ur Experimentalphysik, Universit\"at Innsbruck, Technikerstraße 25/4, 6020 Innsbruck, Austria}

\author{René Nardi \orcidlink{0009-0002-7533-6126}}
\affiliation{Institut f\"ur Experimentalphysik, Universit\"at Innsbruck, Technikerstraße 25/4, 6020 Innsbruck, Austria}

\author{Piotr Gniewek\orcidlink{0000-0002-3869-9342}} \affiliation{Faculty of Physics, University of Warsaw, Pasteura 5, 02-093 Warsaw, Poland}
\author{Micha{\l} Tomza \orcidlink{0000-0003-1792-8043}} 
\affiliation{Faculty of Physics, University of Warsaw, Pasteura 5,
02-093 Warsaw, Poland}

\author{Brandon J. Furey \orcidlink{0000-0001-7535-1874}}
\affiliation{Institut f\"ur Experimentalphysik, Universit\"at Innsbruck, Technikerstraße 25/4, 6020 Innsbruck, Austria}

\author{Philipp Schindler \orcidlink{0000-0002-9461-9650}}
\email{philipp.schindler@uibk.ac.at}
\affiliation{Institut f\"ur Experimentalphysik, Universit\"at Innsbruck, Technikerstraße 25/4, 6020 Innsbruck, Austria}

\date{May 2024}

%\doublespacing

\begin{abstract}
Molecular ions that are generated by chemical reactions with trapped atomic ions can serve as an accessible testbed for developing molecular quantum technologies. On the other hand, they are also a hindrance to scaling up quantum computers based on atomic ions as unavoidable reactions with background gas destroy the information carriers. Here, we investigate the single- and two-photon dissociation processes of single CaOH$^+$ molecular ions co-trapped in Ca$^+$ ion crystals using a femtosecond laser system. We report the photodissociation cross section spectra of CaOH$^+$ for single-photon processes at $\lambda=245 - 275 \,$nm and for two-photon processes at $\lambda=500 - 540\,$nm.
\mt{Measurements are interpreted with quantum-chemical calculations, which predict the photodissociation threshold for $\text{CaOH}^+\to \text{Ca}^++\text{OH}$ at 265\,nm.} 
This result can serve as a basis for dissociation-based spectroscopy for studying the internal structure of CaOH$^+$. 
The result also gives a prescription for recycling Ca$^+$ ions in large-scale trapped Ca$^+$ quantum experiments from undesired CaOH$^+$ ions formed in the presence of background water vapor. 
\end{abstract}

\maketitle

\section{Introduction}
Molecules possess various internal degrees of freedom that are both chemically and physically intriguing. Molecular ions in radio-frequency traps enable precision molecular spectroscopy on the single molecule scale and are thus a promising platform for applications of molecules in quantum technologies\cite{demilleQuantumComputationTrapped2002,linQuantumEntanglementAtom2020} and the exploration of fundamental physics\cite{roussyImprovedBoundElectron2023,alighanbariPreciseTestQuantum2020}. For molecular ions, direct laser cooling via cycling transitions proves to be challenging and has not yet been realized experimentally\cite{nguyenChallengesLasercoolingMolecular2011,ivanovSearchMolecularIons2020,wojcikProspectsOpticalCycling2022}. While the translational motion of trapped molecular ions can be sympathetically cooled by laser cooling co-trapped atomic ions\cite{molhaveFormationTranslationallyCold2000,rugangoSympatheticCoolingMolecular2015,wanEfficientSympatheticMotionalgroundstate2015}, the absence of suitable cycling transitions poses a challenge in investigating the internal structure of molecular ions. Most spectroscopic experiments with trapped molecular ions rely on quantum logic methods utilizing co-trapped atomic qubits\cite{wolfNondestructiveStateDetection2016,chouPreparationCoherentManipulation2017,sinhalQuantumnondemolitionStateDetection2020,chouFrequencycombSpectroscopyPure2020} or destructive detection methods based on, e.g., photodissociation channels\cite{koelemeijVibrationalSpectroscopyMathrm2007,karrPhotodissociationTrappedMathrm2012,seckRotationalStateAnalysis2014,niStatespecificDetectionTrapped2014,khanyileObservationVibrationalOvertones2015,rugangoVibronicSpectroscopySympathetically2016}. 

Various molecular ions can be generated from trapped atomic ions colliding with background gas molecules in a vacuum chamber. Photodissociation of molecular ions generated in this way leads to the recovery of atomic ions. This can be observed from the increase in the overall atomic ion fluorescence\cite{okadaAccelerationChemicalReaction2003,bertelsenRotationalTemperaturePolar2006}, which provides a way to study the rovibrational structure of such molecular ions. 

\begin{figure}
    \centering
    \includegraphics[width=\linewidth]{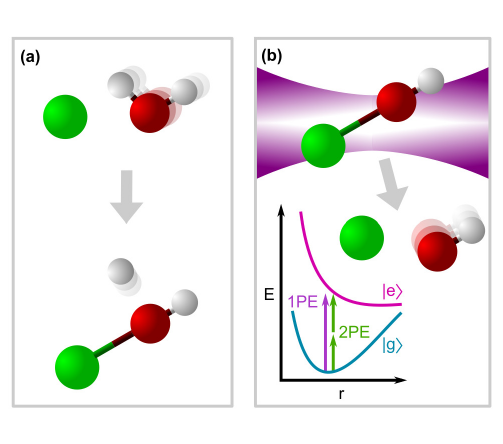}
    \caption{(a) The generation of CaOH$^+$ by a trapped Ca$^+$ ion colliding and reacting with a water molecule. (b) Photodissociation of CaOH$^+$ back to Ca$^+$ by single-photon excitation (1PE) or two-photon excitation (2PE) to its unbound first electronic excited state.}
    \label{fig:photodissociation}
\end{figure}
For trapped ion experiments based on $^{40}$Ca$^+$, which is one of the most well-studied and widely implemented ion species for quantum information processing\cite{schindlerQuantumInformationProcessor2013}, molecular ions generated in chemical reactions with background gas are predominantly CaH$^+$, CaO$^+$, and CaOH$^+$\cite{hansenSingleIonRecyclingReactions2012}. While extensive theoretical and experimental research has been conducted on the two diatomic species\cite{abeInitioStudyPotential2012,rugangoVibronicSpectroscopySympathetically2016,chouPreparationCoherentManipulation2017,vangundySpectroscopyLowlyingStates2018,chouFrequencycombSpectroscopyPure2020,qiAdiabaticallyControlledMotional2023}, there are few spectroscopic studies concerning the triatomic CaOH$^+$ molecular ion. 

As illustrated in Fig.~\ref{fig:photodissociation}.(a), CaOH$^+$ can form in a trapped Ca$^+$ ion experiment in the presence of water vapor in the vacuum chamber by the reaction\cite{okadaAccelerationChemicalReaction2003}
$$
    \text{Ca}^+ + \text{H}_2\text{O} \rightarrow \text{CaOH}^+ + \text{H}.
$$
While CaH$^+$ and CaO$^+$ can be photodissociated with the 397$\,$nm and 375$\,$nm lasers used in trapped Ca$^+$ experiments\cite{wuIncreaseBariumIontrap2021}, the photodissociation of CaOH$^+$ requires a shorter wavelength for exciting the molecule to its unbound first electronic excited state. This photodissociation channel was studied by Okada et al. in 2006 using a combination of optical filters and a broadband UV lamp, where they estimated the dissociation threshold to be $\lambda \approx 255 \, \text{nm}$\cite{okadaPhotodissociationCaOHRegeneration2006}. However, it was later reported by Hansen et al. that Ca$^+$ can also be recovered from CaOH$^+$ by applying 272$\,$nm light\cite{hansenSingleIonRecyclingReactions2012}. In this paper, we report on a measurement of the photodissociation spectrum of CaOH$^+$, where the wavelength-dependence of the photodissociation cross section was measured by excitation with a femtosecond optical parametric amplifier (OPA) system. With the same ultrafast laser system, we are also able to investigate two-photon dissociation of CaOH$^+$ and the measured two-photon dissociation spectrum is reported. The result implies the possibility of recycling Ca$^+$ from CaOH$^+$ with ablation lasers at 515$\,$nm, which are commonly used in trapped Ca$^+$ quantum experiments. \mt{We employ quantum-chemical calculations to reproduce the measured photodissociation energy and predict the spectrum of excited electronic states of \OH.}  

\section{Experimental setup}
\begin{figure*}[htp]
    \centering
    \includegraphics[width=0.9\textwidth]{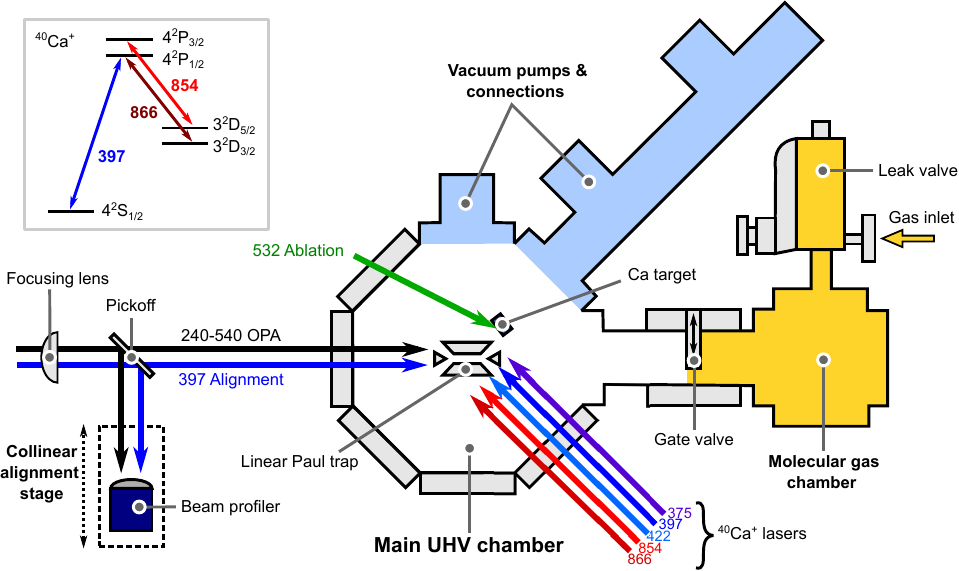}
    \caption{Schematic of the experimental setup and the level structure of $^{40}$Ca$^+$.}
    \label{fig:chamber}
\end{figure*}
As illustrated in Fig.~\ref{fig:chamber}, our experiment is based on a linear Paul trap\cite{pogorelovCompactIonTrapQuantum2021} in ultrahigh vacuum (UHV). A gate valve separates the chamber into two parts: the main experiment chamber and the molecular gas chamber. The main experiment chamber containing the ion trap is connected to a non-evaporable getter (NEG) pump and a NEG-ion combination pump to maintain a base pressure $\sim$ $10^{-10}\,$mbar. Ca$^+$ ions in the ion trap originate from an integrated $^{40}$Ca target mounted near the trap. By ablating the target with 532$\,$nm laser pulses, neutral $^{40}$Ca atoms are introduced to the ion trap where they are photoionized in a two-step process using 375$\,$nm and 422$\,$nm lasers\cite{pogorelovCompactIonTrapQuantum2021}. The generated Ca$^+$ ions are then captured by the ion trap and Doppler-cooled by driving their 4$^2$S$_{1/2}$ $\leftrightarrow$ 4$^2$P$_{1/2}$ cycling transition at 397$\,$nm, as shown in the inset of Fig.~\ref{fig:chamber}. Additional lasers at 866$\,$nm and 854$\,$nm are applied to repump the ions that decay to the 3$^2$D$_{3/2}$ and 3$^2$D$_{5/2}$ states, respectively, back to the cyclic cooling manifold. 

To the right of the main chamber, the molecular gas chamber is used to accumulate water vapor for the generation of CaOH$^+$ molecular ions. When the gate valve is closed, this section is not pumped by the vacuum pumps attached to the main chamber and the water vapor pressure inside increases over time due to outgassing from the chamber wall. To increase the partial pressure of water further, water vapor that is outgassing from an unbaked stainless steel hose is introduced to the molecular gas chamber via a leak valve.

\section{Molecule Generation and identification}
To generate CaOH$^+$ for the photodissociation measurement, water vapor is introduced to the main chamber by opening the gate valve. In our setup, the partial pressure of water vapor in the system cannot be directly measured. Instead, the increase in water vapor pressure in the main chamber is inferred from the change in total pressure estimated from the current of the ion pump. While waiting for the generation of CaOH$^+$, the gate valve is opened every 5 minutes for $\sim$ 1 second, which results in the total pressure estimated by the ion pump increasing from $\sim 10^{-10}\,$mbar to $\sim 10^{-9}\,$mbar. 

As investigated by Okada et al. in 2003, the chemical reaction rate between H$_2$O and the trapped Ca$^+$ ions increases when the ions are pumped into the 3$^2$D$_{3/2}$ state by switching off the 866$\,$nm repumping light during laser cooling\cite{okadaAccelerationChemicalReaction2003}. Therefore, the 866$\,$nm laser is periodically switched off for 1 second every 2 seconds in the molecule generation phase. The generation rate of one CaOH$^+$ molecular ion is typically 10 to 15 minutes for the ion crystals of $3-8$ Ca$^+$ ions investigated in this experiment.

\begin{figure} 
    \centering
    \includegraphics{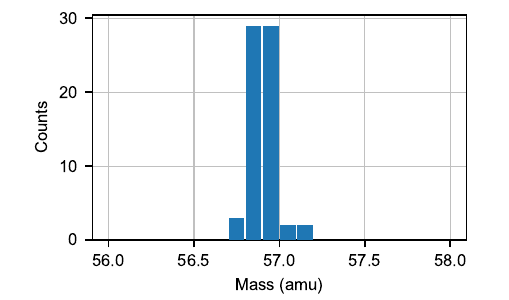}
    \caption{Results of 65 mass spectrometry measurements of the generated molecular ion species distribution.}
    \label{fig:mass-spec}
    %add inset
\end{figure}

The above molecule generation process was found to produce predominantly CaOH$^+$ by mass spectrometry using the method presented by Drewsen et al. \cite{drewsenNondestructiveIdentificationCold2004} (for details, refer to Appendix~\ref{apd:mass spec}). In this method, the unknown molecular mass $m_{\textrm{x}}$ can be inferred from the frequency $\omega_x$ of the axial center-of-mass (COM) motion of a 2-ion crystal of one Ca$^+$ ion and one molecular ion\cite{kielpinskiSympatheticCoolingTrapped2000,drewsenNondestructiveIdentificationCold2004} as
\begin{equation}
\label{Equ_Ref_mass_spec}
    \omega_x^2 = \big( 1 + \xi - \sqrt{1 - \xi + \xi^2} \big)\, \omega_0^2,
\end{equation}
where $\xi=m_0/m_x$ is the ratio between the atomic and molecular mass and $\omega_0$ is the COM motional frequency of a Ca$^+$ ion crystal. These frequencies were measured by applying an oscillating voltage to one of the endcap electrodes. When the frequency of this ``tickling'' signal is resonant with that of the COM motion of the ion crystal, it excites the corresponding oscillation of the crystal. This increases the motional temperature of the ion crystal and thus reduces the scattered Doppler cooling light. The COM motional frequency is thus determined from the ion crystal's fluorescence as a function of the tickling frequency. The masses of the generated molecular ions are measured with a precision of $\delta$m $\approx 0.1\,$amu and the results are shown in Fig.~\ref{fig:mass-spec}. All 65 measurements have masses consistent with CaOH$^+$, from which we conclude that the molecular ions generated in our procedure are predominantly this species.

\section{Photodissociation}
After verifying the generation of a single CaOH$^+$, the measurement of the photodissociation spectrum of CaOH$^+$ is performed by applying femtosecond pulse trains produced by an OPA. The spatial beam profile and position of the OPA light are measured by a beam profiler mounted on a rail. To assist the alignment of the OPA light, an additional 397$\,$nm Doppler cooling beam is applied co-propagating with the OPA light, which is aligned to the center of the ion trap by maximizing the 397$\,$nm light scattered by the ion crystal. The OPA light is overlapped with this axial Doppler cooling beam at the position of the ions. We estimate the alignment precision of this procedure to be 10$\,$µm, limited by the resolution of the beam profiler. The radius of the OPA light at the ion crystal was also estimated by the beam profiler to be 150(10)$\,$µm. 

We assume that the dissociation of a CaOH$^+$ molecule during illumination is memoryless due to the lack of intermediate electronic states. Thus, the dissociation probability after illumination time $t$ can be described by the cumulative distribution function (CDF) of an exponential distribution,
\begin{equation}
    \label{eq:ex dis}
        P(t;T_\textrm{D}(\lambda)) = 1 - e^{ - t/T_\textrm{D}(\lambda)},
\end{equation}
characterized by the mean dissociation time $T_\textrm{D}(\lambda)$, where $\lambda$ is the wavelength of the OPA light. The effective photodissociation cross section can be related to the mean dissociation time by (see Appendix~\ref{apd:1pd cs})
\begin{equation}
    \label{1photon cs}
    \sigma_{\textrm{eff}} = \frac{\epsilon_{\lambda}}{T_\textrm{D}(\lambda) I_{\textrm{avg}}}.
\end{equation}
Here, $\epsilon_{\lambda}=h c / \lambda$ is the photon energy, $h$ is the Planck's constant, $c$ is the speed of light, and $I_\mathrm{avg} = 2 P_{\textrm{avg}}/(\pi w^2)$ is the average intensity of the OPA light at the ion position with average power $P_{\textrm{avg}}$ and beam radius $w$.

\begin{figure*}
    \centering
    \includegraphics[width=\linewidth]{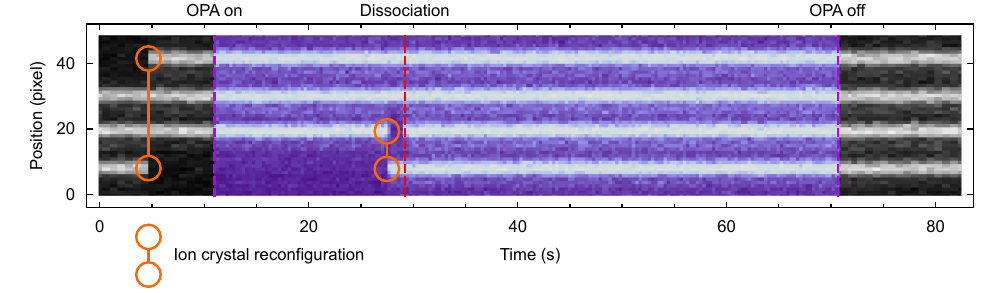}
    \caption{Visualization of the time evolution of the ion crystal configuration using a series of fluorescence images of a typical photodissociation trial. The images are acquired over an exposure time of 0.2$\,$s and are integrated perpendicular to the trap axis to obtain the positions of bright Ca$^+$ ions and dark CaOH$^+$ ions.}
    \label{fig:y-lines}
\end{figure*}

The experimental procedure is initialized by loading a one-dimensional ion chain with $3 - 8$  Ca$^+$ ions into the ion trap for the generation of a single CaOH$^+$ (see Fig.~\ref{fig:exp sch}). While Ca$^+$ ions can be detected on an electron-multipling CCD (EMCCD) camera, the generated CaOH$^+$ ions do not scatter light and are recognized as dark defects in the ion chain. The number of dark ions is identified by comparing the ion crystal image with pre-calculated ion crystal configuration templates with different numbers of dark and bright ions (see Appendix~\ref{apd:ion finder}). After the appearance of a single molecular ion, the OPA light is applied to illuminate the molecular ion for up to 300$\,$s. If the OPA light successfully recovers the original Ca$^+$ from the CaOH$^+$ ion, as shown in Fig.~\ref{fig:y-lines}, the illumination time for achieving photodissociation will be recorded. Otherwise, if the photodissociation is not successful or the dark ion is lost during the process, the dissociation trial is marked as unsuccessful within the accumulated illumination time. In either case, utilizing Bayesian analysis, the $n$-th trial provides information to update the prior distribution of the mean dissociation time $T_\textrm{D}$ by multiplying it with a likelihood function 
\begin{equation}
\mathcal{L}_{n}(T_\textrm{D}=T) = 
\begin{cases}
\frac{1}{T} e^{ - T_n/T} & \text{success at }T_n \\
e^{ - T_{\textrm{total}}/T} & \text{fail}
\end{cases}
\end{equation}
based on Eq.~(\ref{eq:ex dis}) where $T_{\textrm{total}}$ denotes the total illumination time.
Starting from a uniform prior distribution in the range of $0 - 3000\,$s, the posterior distribution after $N$ trials,
\begin{equation}
\mathcal{L}_{\textrm{post}}(T_\textrm{D}=T) = \overset{N}{\prod_{n=1}} \mathcal{L}_{n}(T_\textrm{D}=T),
\end{equation}
is normalized and reveals an estimation of $T_\textrm{D}$. To ensure convergence of the posterior distribution, approximately 20 trials were performed at each wavelength.

\begin{figure}
\centering
\includegraphics{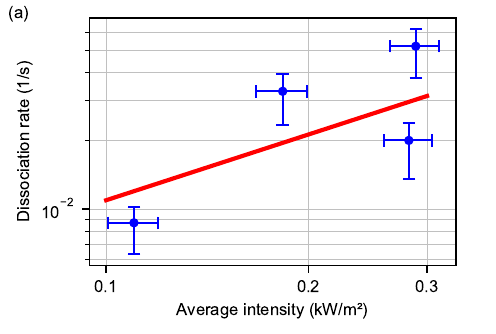}
\hfill
\centering
\includegraphics{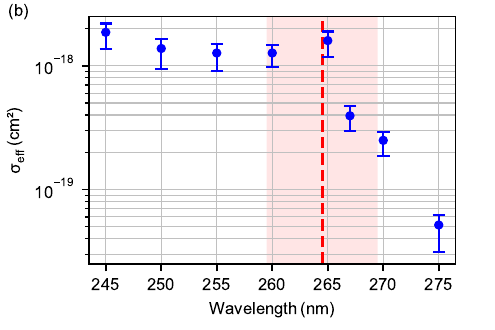}
\caption{(a) The measured dissociation rate vs. the average intensity of the OPA light at 250$\,$nm. The red line is a power law fit of the data with the exponent $\alpha=0.9^{+0.8}_{-0.3}$. (b) The photodissociation spectrum of CaOH$^+$ at $\lambda=245 - 275\,$nm with the effective photodissociation cross section calculated from Eq.~(\ref{1photon cs}). Theory prediction for photodissociation threshold at $\lambda=264.5\,$nm is given as a red dashed line, around which a shaded area of $\pm 5\,$nm shows the conservative estimation of the theory uncertainty.}
\label{fig:1photon}
\end{figure}

As shown in Eq.~(\ref{1photon cs}), linear dependence of the dissociation rate on the average power of the OPA light is expected. This is verified in the measurement of the dissociation rate at different $I_{\textrm{avg}}$ at $\lambda=250\,$nm. Fitting the results to a power law function, as shown in Fig.~\ref{fig:1photon}.(a), reveals an exponent of $\alpha=0.9^{+0.8}_{-0.3}$. Thus, to better estimate the mean dissociation time, the average power of the beam is adjusted between 6 and 60$\,$µW in order for at least half of the dissociation trials to be successful within the maximum illumination time. The repetition rate of the OPA is fixed at 1$\,$kHz. The photodissociation cross sections at wavelengths ranging from 245 to 275$\,$nm are then measured and shown in Fig.~\ref{fig:1photon}.(b).

\section{Two-photon dissociation}

\begin{figure}
\centering
\includegraphics{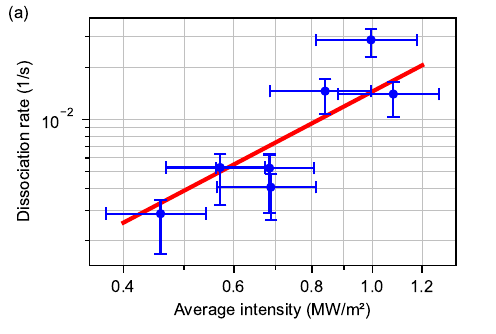}
\hfill
\includegraphics{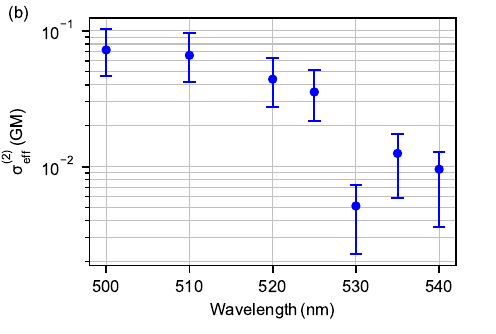}
\caption{(a) The measured photodissociation rate vs. average intensity of the OPA light at $\lambda=500\,$nm. The red line is a power law fit of the data with exponent $\alpha=1.9^{+1.2}_{-0.8}$. 
(b) The two-photon dissociation spectrum of CaOH$^+$ at $\lambda=500 - 540 \,$nm with the two-photon dissociation cross sections calculated from Eq.~(\ref{2photon cross section}).}
\label{fig:2photon}
\end{figure}

In addition to the measurement of single-photon dissociation, we also investigated the photodissociation of CaOH$^+$ at wavelengths around 500$\,$nm. To achieve a dissociation rate that is comparable to the single-photon measurements, the repetition rate of the OPA is increased to 10$\,$kHz and the light intensity is increased by reducing the beam radius to 70(10)$\,$µm and increasing the average power to a few milliwatts. At $\lambda=500\,$nm, the dependence of the dissociation rate on the average intensity of the OPA light on the ions is investigated, as illustrated in Fig.~\ref{fig:2photon}.(a). Fitting the data to a power law function gives an exponent of $\alpha=1.9^{+1.2}_{-0.8}$, which is consistent with two-photon processes.

Thus, the photodissociation of CaOH$^+$ at wavelengths around 500$\,$nm is assumed to be dominated by a two-photon dissociation channel via its unbound first electronic excited state. The two-photon dissociation cross section $\sigma^{(2)}_{\textrm{eff}}$ is related to the dissociation rate by (see Appendix~\ref{apd:2pd cs})
\begin{equation}
    \frac{1}{T_\textrm{D}} = f_{\textrm{rep}} \int_{\textrm{pulse}}{\frac{\sigma^{(2)}_{\textrm{eff}}}{\epsilon_{\lambda}} \frac{I(t)^2}{2\epsilon_{\lambda}} \mathrm{d}t},
\end{equation}
which gives
\begin{equation} \label{2photon cross section}
    \sigma^{(2)}_{\textrm{eff}} = \sqrt{\frac{2 
\, \textrm{ln}2}{\pi}} \frac{4 \epsilon_{\lambda}^2}{T_\textrm{D} f_{\textrm{rep}} \tau_\mathrm{p} I_{0}^2},
\end{equation}
in units of GM (1$\,$GM = $10^{-50} \, \textrm{cm}^4 \, \textrm{s} \, \textrm{photon}^{-1}$), where a Gaussian temporal pulse profile  $I(t)=I_0 \, 2^{- (2t)^2/\tau_\mathrm{p}^2}$ is assumed and $\tau_\mathrm{p}$ is the full width at half maximum (FWHM) pulse duration. The peak intensity $I_0$ is related to the average intensity by $I_{\textrm{avg}} = \sqrt{\pi /4 \,
\textrm{ln}2}\, I_0 \tau_\textrm{p} f_{\textrm{rep}}$. The pulse duration at each wavelength is assumed to be Fourier-transform-limited and is thus estimated from the spectral bandwidth of the OPA light (see Appendix~\ref{apd:opa}). The two-photon dissociation cross section spectrum at $\lambda=500 - 540\,$nm is then measured and shown in Fig.~\ref{fig:2photon}.(b).

\section{Quantum-chemical modelling}

\begin{figure}
\centering
\includegraphics[width=\columnwidth]{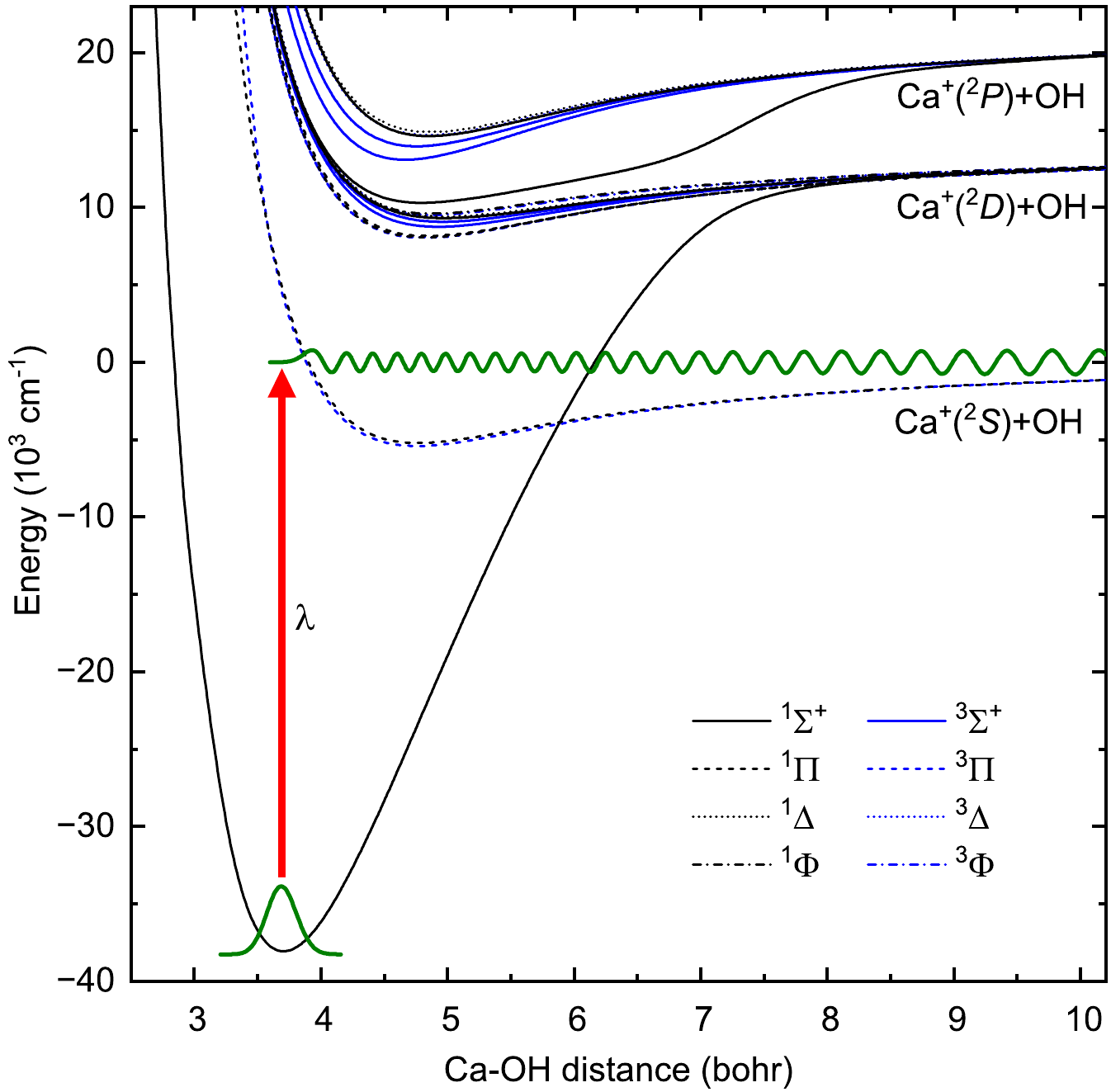}
\caption{One-dimentional cut through the potential energy surfaces for the ground and excited electronic states of CaOH$^+$ in the linear geometry as a function of the distance between Ca$^+$ and the center of mass of OH. The arrow indicates the photodissociation transition. Nuclear wave functions for the ground-state stretching of CaO bond and zero-energy scattering of Ca$^+$ + OH are plotted.}
\label{fig:PESs}
\end{figure}

\mt{To interpret experimental measurements and predict theoretical spectroscopic parameters of CaOH$^+$, we use state-of-the-art quantum-chemical methods. We employ wave-function approaches: the coupled cluster method restricted to single, double, and noniterative triple excitations, CCSD(T)~\cite{HampelCPL92}, to describe the ground electronic state and internally contracted multireference configuration interaction method restricted to single and double excitations, MRCISD~\cite{WernerJCP88}, to describe excited electronic states. Electronic orbitals are constructed using the augmented correlation-consistent polarized weighted core-valence quintuple-$\zeta$ quality basis set (aug-cc-pwCV5Z-PP~\cite{HillJCP17} for Ca, aug-cc-pwCV5Z~\cite{PetersonJCP02} for O, and aug-cc-pV5Z~\cite{DunningJCP89} for H). We include the scalar relativistic effect in Ca using the small-core relativistic energy-consistent pseudopotential (ECP10MDF~\cite{LimJCP06}) to replace the inner-shell electrons. Electronic structure calculations are performed with the \textsc{Molpro}~\cite{Molpro,MOLPRO-WIREs} package of \textit{ab initio} programs. Uncertainties of present energy calculations should not be larger than 1-2\%\cite{GronowskiPRA20}.}

\mt{We optimize the ground-state geometry of linear CaOH$^+$ and find the equilibrium internuclear distances $R_e^\text{CaO}=3.574\,$bohr and $R_e^\text{OH}=1.803\,$bohr. Such an equilibrium geometry gives the rotational constant $B_e=0.3655\,$cm$^{-1}$. Present theoretical harmonic frequencies $\omega_e$ for normal modes of CaOH$^+$ are 3949.9$\,$cm$\,^{-1}$ for OH stretching, 747.4$\,$cm$^{-1}$ for CaO stretching, and 481.2$\,$cm$^{-1}$ for doubly degenerate bending. The frequency for the OH mode is slightly larger as compared to our theoretical value of 3748$\,$cm$^{-1}$ and the experimental value of 3737.8$\,$cm$^{-1}$~\cite{Huber1979} for a free OH. Additionally, using the finite-field approach, we predict the equilibrium permanent electric dipole moment of CaOH$^+$ with respect to its center of mass to be 6.3 debye. This relatively large dipole moment originates from the partially charge-transferred nature of the ground-state CaOH$^+$, which can be formally described as Ca$^{2+}$OH$^-$.}

Under room temperature (300K), the molecular ion is likely to be in its vibrational ground state (97\% for CaO stretching).
\mt{Using the CCSD(T) method, we predict that the dissociation energy of CaOH$^+$ in the ground rovibrational level ($v=0,j=0$) into $\text{Ca}^+$ and OH ($v=0,j=0$) is 37807$\,$cm$^{-1}$, which corresponds to 264.5$\,$nm. This value originates from 38763$\,$cm$^{-1}$ of the electronic
contribution corrected by $-956\,$cm$^{-1}$ of the zero point energy (ZPE) difference within the harmonic approximation, where the ZPE for CaOH$^+$ and OH is 2830$\,$cm$^{-1}$ and 1874$\,$cm$^{-1}$, respectively.}

\mt{Figure~\ref{fig:PESs} presents the ground and excited electronic states of the CaOH$^+$ in the linear geometry as a function of the distance between Ca and the center of mass of OH. OH is described within the rigid-rotor approximation with the experimental internuclear distance of $r_\text{OH}$ = 1.8332$\,$bohr~\cite{DrouinJPCA13}. The one-photon dissociation of CaOH$^+$ is driven by the allowed electric dipole transition from the ground $^1\Sigma^+$ state to the excited $^1\Pi$ state, which is slightly repulsive around the ground-state equilibrium geometry and dissociates into Ca$^+$ and OH. The corresponding vertical transition moment obtained with the MRCISD method is 0.8$\,$debye. The photodissociation cross sections are given by the convolution of the transition moment function and the overlap of the ground rovibrational and excited continuum nuclear wave functions. Exact values of cross sections are sensitive to the precise location of the classical turning point of the $^1\Pi$ state, which is, however, robustly predicted by the employed \textit{ab initio} methods to be within the width of the ground rovibrational wave function.}

\section{Discussion}
In summary, we measured the photodissociation spectrum of single trapped CaOH$^+$ molecular ions sympathetically cooled in Ca$^+$ ion crystals at $\lambda = 245 - 275\,$nm and $\lambda = 500 - 540\,$nm. The photodissociation processes at wavelengths in these two ranges are verified to originate predominantly from a single-photon and a two-photon dissociation channel, respectively. While the one-photon dissociation cross section started to drop off around $\lambda=265\,$nm, a similar decrease was observed in the two-photon dissociation spectrum around $\lambda=525\,$nm. This implies that both processes involve the excitation of the molecular ion to its unbound first electronic excited state which leads to its dissociation. Compared to the previous work\cite{okadaPhotodissociationCaOHRegeneration2006} on single-photon dissociation of CaOH$^+$ by Okada et al. in 2006, we report a more precise measurement of the dissociation cross sections at different wavelengths which is consistent with the previous estimation of $\sigma_{\textrm{eff}} \approx 10^{-18} \textrm{cm}^2$ at $\lambda=255\,$nm. The measured single-photon dissociation spectrum is also consistent with the photodissociation of CaOH$^+$ at 272$\,$nm observed by Hansen et al.\cite{hansenSingleIonRecyclingReactions2012}.

As for the two-photon dissociation of CaOH$^+$ around $\lambda=500\,$nm, the dissociation process is found to be significantly less efficient compared to the single-photon dissociation process, as expected since higher peak intensity is typically required for higher-order processes. Yet femtosecond lasers at such wavelengths have proven to be alternatives to continuous wave lasers with $\lambda<265\,$nm for photodissociation of the molecular ion. The question arises as to whether nanosecond lasers that are used for ablation loading in trapped atomic ion experiments at $\lambda = 515\,$nm\cite{pogorelovCompactIonTrapQuantum2021} are able to dissociate CaOH$^+$. Achieving a mean photodissociation time of 1 minute for a single CaOH$^+$ ion is estimated to require an average power of $0.2 - 0.5\,$W, considering a laser with $f_{\textrm{rep}} = 2\,$kHz, $\tau_\textrm{p} = 1\,$ns, and the light focused onto the ions with a beam waist of $w = 60 \,$µm. 

\mt{We employed advanced \textit{ab initio} molecular electronic structure methods to characterize theoretically the ground-state properties of CaOH$^+$ and to investigate the spectrum of excited electronic states involved in the measured photodissociation. We predicted the photodissociation threshold for CaOH$^+$$\to$Ca$^+$+OH at 265$\,$nm in very good agreement with the one-photon experimental observations. Efficient photodissociation is possible because of the coincidence of the slightly repulsive part of the excited $^1\Pi$ electronic state close to its inner classical turning point, which is accessible by a decent electric dipole transition moment.}

The reported measurement is implemented on single CaOH$^+$ molecular ions in small trapped ion crystals, which enables the relatively precise evaluation of the photodissociation cross section due to the repeatable positioning of the molecule. The scale of the system is also crucial for investigating a two-photon dissociation process that requires tightly focused light to achieve sufficient intensity. The reported photodissociation spectrum can be a basis for studying the rovibrational structure of CaOH$^+$ via dissociation-based spectroscopy. Designed to investigate the interaction between femtosecond laser pulses and a linear ion crystal with a single CaOH$^+$, this experiment represents a precursor for performing quantum logic spectroscopy with polyatomic molecular ions \cite{schindlerUltrafastInfraredSpectroscopy2019}.

\section*{Acknowledgements}
This research was funded by ERC-2020-STG 948893, ESQ Discovery project SDEF: State-dependent force spectroscopy for trapped ions, FWF 1000 Ideas project TAI-798, and \mt{the Foundation for Polish Science within the First Team programme}. \mt{We acknowledge Poland’s high-performance computing infrastructure PLGrid (HPC Center: ACK Cyfronet AGH) for providing computer facilities and support within computational project PLG/2023/016115.} The authors would like to acknowledge AQT for support with the ion trap and vacuum pumps, in particular Daniel Nigg and Georg Jacob; Markus Teller for assistance with vacuum work; electronics workshop technicians Kilian Prokop and Wolfgang Kuen for lab infrastructure support; Milan Oncak for discussions on quantum chemistry calculations; Christian Marciniak for general discussions; Marco Valentini for Ca$^+$ lasers implementation; and the Quantum Optics \& Spectroscopy Group and associated ion trapping groups at the Universität Innsbruck for general assistance. 

\textit{\textbf{Author Contributions}}  Z.W., S.W., B.F., and P.S. designed the experiment. Z.W., S.W., and B.F. carried out the dissociation experiment and analyzed the data. M.I.M., E.M., and B.F. performed the characterization of the OPA system. \mt{P.G. and M.T. performed electronic structure calculations.} Z.W., S.W., B.F. and M.T. contributed to the manuscript. Z.W., S.W., V.P., M.I.M., E.M., G.M., R.N., B.F., and P.S. contributed to the experimental setup. All authors reviewed the manuscript. B.F. and P.S. supervised the project.

\section*{Conflict of Interest}
The authors have no conflicts of interest to disclose.

\section*{Data and Code Availability}
The data that support the findings of this study are openly available at \href{http://doi.org/10.5281/zenodo.11109790}{http://doi.org/10.5281/zenodo.11109790}.

\appendix

\section{Photodissociation model}
\subsection{Single-photon dissociation}\label{apd:1pd cs}
The photodissociation process considered in this paper is treated as a consequence of a photon absorption event which excites the molecular ion to its unbound first electronic excited state. We thus define the effective photodissociation cross section $\sigma_{\mathrm{eff}} = \eta \sigma$ in terms of the absorption cross section $\sigma$ and the quantum yield $\eta$. We assume all of the photon absorption events lead to dissociation of the molecular ion, i.e., $\eta = 1$, and the molecular rovibrational structure is not considered for simplicity.

The absorption cross section of a single molecule is related to the absorption coefficient $\alpha$ of molecular gas by 
\begin{equation}
    \label{alpha 1photon}
    \alpha = n \, \sigma,
\end{equation}
where $n$ denotes the molecular number density. For light propagating through the absorptive molecular gas, the change in light intensity is described by the differential equation
\begin{equation}
    \label{eq:diff 1photon}
    \frac{\mathrm{d}I}{\mathrm{d}z} = - \alpha I.
\end{equation}
Hereby we consider photons absorbed by molecules in a small fraction of the molecular gas with volume $\Delta V = \Delta A \Delta z$ and the photon absorption event in each molecule that leads to the excitation of the molecule is considered as independent. Regarding a train of ultrafast laser pulses separated in time by $f_{\textrm{rep}}^{-1}$, the mean number of excitations induced by each laser pulse can be expressed as
\begin{equation}
    \label{eq: mean ex per pulse}
    \delta \langle N_{\textrm{ex}} \rangle = \int_{\textrm{pulse}} \frac{\alpha I \Delta A \Delta z}{\epsilon_{\lambda}} dt = \frac{\sigma \Delta N}{\epsilon_{\lambda}} \int_{\textrm{pulse}} I \mathrm{d}t,
\end{equation}
where $\epsilon_{\lambda}$ is the photon energy and $\Delta N = n \Delta V$ is the number of molecules.

As the measured dissociation time $T_{n} \gg f_{\textrm{rep}}^{-1}$ is at a larger time scale, the mean number of excitations per unit time is given as
\begin{equation}
    \label{eq: mean ex per unit time}
    \frac{\mathrm{d} \langle N_{\textrm{ex}} \rangle}{\mathrm{d}T} = \frac{\mathrm{d} N_{\textrm{pulse}}}{\mathrm{d}T} \delta \langle N_{\textrm{ex}} \rangle = \frac{I_{\textrm{avg}} \sigma}{\epsilon_{\lambda}} \Delta N,    
\end{equation}
where $I_{\textrm{avg}} = f_{\textrm{rep}} \int_{\textrm{pulse}}I \, \mathrm{d}t$ and $f_{\textrm{rep}} = \mathrm{d} N_{\textrm{pulse}}/ \mathrm{d}T$. Considering a single molecule ($\Delta N=1$) that can only be excited once, the probability of the molecule being excited per unit time is
\begin{equation}
\frac{\mathrm{d}P_{\textrm{ex}}(T)}{\mathrm{d}T} = \left[ 1-P_{\textrm{ex}}(T) \right] \frac{I_{\textrm{avg}} \sigma}{\epsilon_{\lambda}} ,     
\end{equation}
where $P_{\textrm{ex}}(T)$ is then the exponential distribution CDF, 
\begin{equation}
    P_{\textrm{ex}}(T) = 1 - e^{\sigma I_{\textrm{avg}} T / \epsilon_{\lambda}}.
\end{equation}
Therefore, the mean dissociation time $T_\textrm{D}$ can be related to the absorption cross section and thus the effective dissociation cross section by
\begin{equation}
    \label{eq:mean diss time}
    T_\textrm{D} = \frac{\epsilon_{\lambda}}{\sigma I_{\textrm{avg}}} = \frac{\epsilon_{\lambda}} {\sigma_{\textrm{eff}} I_{\textrm{avg}}}.
\end{equation}

\subsection{Two-photon dissociation}\label{apd:2pd cs}
Similarly, we define the effective two-photon dissociation cross section to be $\sigma_{\textrm{eff}}^{(2)} = \eta \sigma^{(2)}$, related to the two-photon absorption cross section and the quantum yield $\eta = 1$. Considering two-photon absorption in a molecular gas, the change in light intensity is described by
\begin{equation}
    \label{eq:diff 2photon}
    \frac{\mathrm{d} I}{\mathrm{d} z} = - \beta I^2,
\end{equation}
characterized by the two-photon absorption coefficient $\beta=n \sigma^{(2)}/ \epsilon_{\lambda}$, where $\sigma^{(2)}$ is the two-photon absorption cross section in units of GM. Compared to single-photon dissociation, a two-photon dissociation event requires the absorption of two photons to excite the molecule to its unbound excited state. Similar to Eq.~(\ref{eq: mean ex per pulse}), for a train of ultrafast laser pulses separated in time by $f_{\textrm{ref}}^{-1}$, the mean number of excitations induced by each laser pulse can be expressed as
\begin{equation}
    \label{eq: mean ex per pulse 2photon}
    \delta \langle N_{\textrm{ex}} \rangle = \frac{\sigma^{(2)} \Delta N}{2 \epsilon_{\lambda}^2} \int_{\textrm{pulse}} I^2 \mathrm{d}t.
\end{equation}
Assuming the pulses have a Gaussian temporal profile, i.e., $I = I_0\ 2^{-(2 t)^2/\tau_\mathrm{p}^2}$, the mean number of excitations per unit time can be derived as
\begin{equation}
     \frac{\mathrm{d} \langle N_{\textrm{ex}} \rangle}{\mathrm{d}T} = \sqrt{\frac{\pi}{2 \, \textrm{ln} 2}} \frac{\sigma^{(2)} I_0^2 \tau_\mathrm{p} f_{\textrm{rep}}}{4 \epsilon_{\lambda}^2} \Delta N,
\end{equation}
where $\tau_\mathrm{p}$ is the FWHM pulse duration. Similar to the single-photon case in Eq.~(\ref{eq:mean diss time}), the mean dissociation time of a single molecule is thus related to the two-photon dissociation cross section by
\begin{equation}
    T_\textrm{D} = \sqrt{\frac{2 \, \textrm{ln} 2}{\pi}} \frac{4 \epsilon_{\lambda}^2} {\sigma^{(2)}_{\textrm{eff}} \tau_\mathrm{p} f_{\textrm{rep}} I_0^2}.
\end{equation}

\subsection{Memory of the photodissociation process}\label{apd:mln}
To verify the memorylessness of the photodissociation process, the illumination for photodissociation is divided into several time intervals with various fixed durations. From this data, we can estimate the probability of dissociation for a fixed illumination duration as well as the cumulative time until dissociation. If the photodissociation process has no memory effect, the probability of the molecule being dissociated after being illuminated for a fixed time $T$ will be described by the exponential distribution CDF as in Eq.~(\ref{eq:ex dis}). 

\begin{figure}
    \centering
    \includegraphics{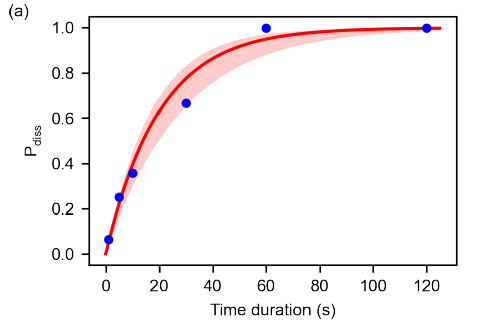}
    \hfill
    \includegraphics{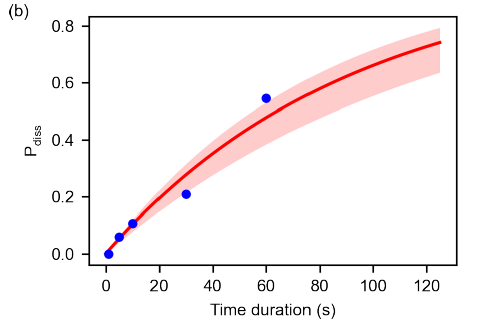}
    \caption{Dissociation probability after applying OPA light for different fixed time durations at (a) $ \lambda=270\,$nm and (b) $\lambda=500\,$nm. Based on Bayesian analysis, the results are fitted to the exponential distribution CDF, as shown by the red curves with the confidence intervals given in light red.}
    \label{fig:interval exp fit plot}
\end{figure}

\begin{figure}
    \centering
    \includegraphics{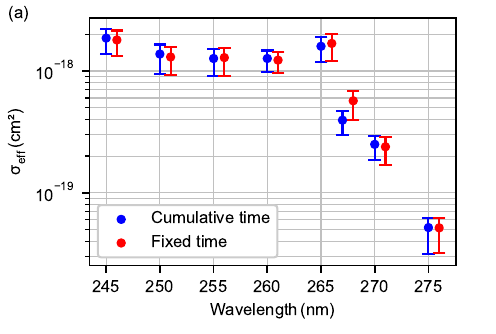}
    \hfill
    \includegraphics{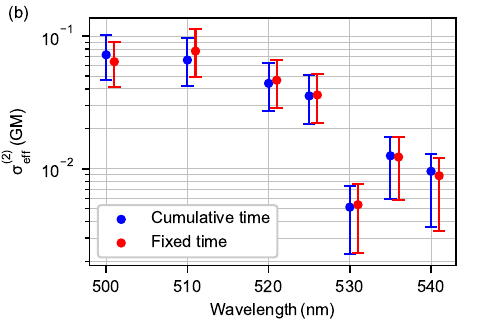}
    \caption{The photodissociation cross section for the (a) single-photon dissociation process and the (b) two-photon dissociation process, as calculated from the mean dissociation time obtained either from the successful dissociation rate after applying OPA light for various time durations (red, shifted right by 1$\,$nm for readability) or from the exact dissociation time for achieving dissociation (blue).}
    \label{fig:cs compare}
\end{figure}

We can therefore also estimate the mean dissociation time using the estimated dissociation probability for fixed times. These probabilities are fitted to the exponential distribution CDF as indicated in Fig.~\ref{fig:interval exp fit plot}, e.g., for $\lambda=270\,$nm and $\lambda=500\,$nm. We estimate the photodissociation cross sections via this analysis for all data and compare them to the cumulative dissociation time, as shown in Fig.~\ref{fig:cs compare}. We find that the cross sections obtained from both methods agree within the uncertainties which indicates that the single- and two-photon dissociation processes are memoryless. 

\section{Experimental details}

\subsection{Ca$^+$ Lasers and trap parameters}
During the photodissociation measurement, the cooling laser at 397$\,$nm and the repumpers at 866$\,$nm and 854$\,$nm are constantly applied for Doppler cooling the ion crystal. The upper limit of the power applied for each laser is 0.6$\,$mW (397$\,$nm), 0.4$\,$mW (866$\,$nm), and 0.3$\,$mW (854$\,$nm). The beam radii of the lasers at the ion position are larger than $50\,$µm. To provide axial confinement of the ion crystal, the voltage on the endcap electrodes of the ion trap is set to 205$\,$V. This leads to an axial COM motional frequency of $\omega_0 = 2\pi \times 526.8(2) \,$kHz for all the experiments in this paper, based on the measurement discussed in Appendix~\ref{apd:mass spec}. 

\subsection{Autocorrelation and spectra of OPA} \label{apd:opa}
\begin{figure}
\centering
\includegraphics{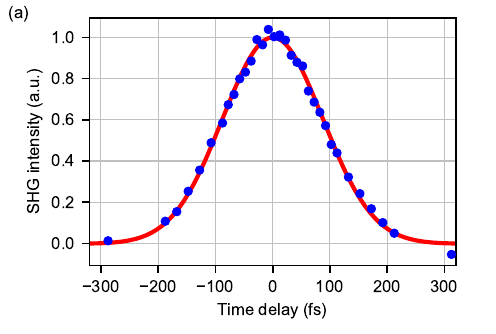}
\hfill
\centering
\includegraphics{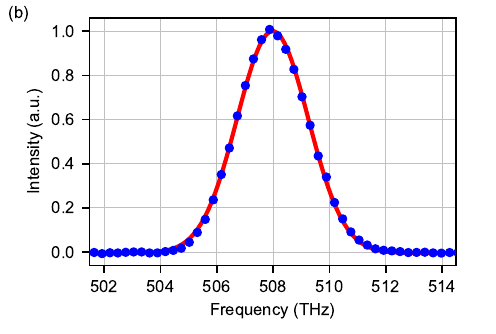}
\caption{(a) The SHG intensity vs. time delay for the autocorrelation performed at $\lambda = 590\,$nm.
(b) The spectral intensity vs. frequency measured at $\lambda = 590\,$nm. The Gaussian fits to the data for both plots are indicated in red.}
\label{fig:acspec}
\end{figure}

The femtosecond OPA system used in this work is an ORPHEUS-HP OPA seeded and pumped by a Carbide CB5 amplifier from Light Conversion. As knowledge of the pulse duration is required to obtain the two-photon dissociation cross section, we characterized it using a combination of measurements of the spectra and autocorrelation measurements. While the pulse duration at $\lambda=500-540\,$nm could not be directly measured with our current setup, the pulses produced by the OPA are guaranteed to be transform-limited by the manufacturer. To verify this, the pulse duration and the spectrum of the OPA at $\lambda = 590\,$nm were measured and the time-bandwidth product was inferred, as shown in Fig.~\ref{fig:acspec}.

To measure the pulse duration, we used an autocorrelator setup that splits the light from the OPA into two paths, where one path has an adjustable time delay. Overlapping both branches on a phase-matched BBO crystal induces type I second-harmonic generation. Assuming a Gaussian temporal profile, the FWHM of the SHG signal was measured to be $\tilde{\tau}_\mathrm{ac} = 206(4)\,$fs, corresponding to a pulse duration of $\tau_\mathrm{p} = \tilde{\tau}_\mathrm{ac}/\sqrt{2} = 146(3)\,$fs. 

The spectra of the OPA light were measured with an Ocean Insight Ocean HR spectrometer. Fitting the spectrum at $\lambda = 590\,$nm to a Gaussian distribution, the FWHM spectral width was determined to be $\Delta \nu = 2.973(6)\,$THz. This gives a time-bandwidth product of $\mathrm{TBP} = \tau_\mathrm{p}\,\Delta \nu = 0.433(8)$, which is in agreement with the time-bandwidth product of a transform-limited Gaussian pulse, i.e., $\mathrm{TBP}_\mathrm{Gaussian} = 0.441$. 

Thus, we assume the pulses are transform-limited at $\lambda=500-540\,$nm in the two-photon dissociation experiments. The measured center wavelengths, their corresponding spectral bandwidths, and the inferred pulse durations are reported in Table~\ref{tab:pulsespectra}.

\begin{table}   
\caption{\label{tab:pulsespectra}Measured spectral bandwidths and inferred pulse durations assuming $\mathrm{TBP} = 0.441$.}
\begin{ruledtabular}
\begin{tabular}{lll}
\mbox{$\lambda\,$(nm)}&\mbox{$\Delta \nu\,$(THz)}&\mbox{$\tau_\mathrm{p}\,$(fs)} \\
\hline
501.63(1) & 3.83(1) & 96.8(3) \\ 
510.16(1) & 4.20(2) & 91.2(4) \\ 
520.526(6) & 4.096(7) & 97.4(2) \\ 
527.339(8) & 3.47(1) & 118.0(3) \\ 
530.778(5) & 2.963(6) & 140.0(3) \\ 
536.025(7) & 3.489(9) & 121.3(3) \\ 
540.597(8) & 3.06(1) & 140.6(4) \\ 
\end{tabular}
\end{ruledtabular}
\end{table}

\subsection{Error estimation}
In addition to the statistical uncertainties in the measured parameters of OPA pulse duration, the following sources of uncertainty are considered. The uncertainties of the mean dissociation times are given by Bayesian analysis, which are significantly larger than the uncertainties of the measured dissociation times. The uncertainty of the average intensity of the OPA light at the ion crystals derives from the uncertainty in average power ($0.1\,$µW in single-photon measurements, $0.1\,$mW in two-photon measurements), the uncertainty of beam positioning of ($10\,$µm), and the uncertainty of the beam radius. The beam radius is estimated by the beam profiler with a precision of $10\,$µm at a position corresponding to the real ion crystal position in the beam path. The deviation from the real ion crystal position is $< 5\,$mm and results in a possible systematic error in the measured beam radius at the ion crystal, which is $<10\,$µm. The total uncertainty of the beam radius is thus considered to be $10\,$µm.

\subsection{Confidence}
The uncertainties presented in this paper are based on a 68\% confidence level.

\subsection{Photodissociation experiment process tree}
\begin{figure}
\centering
\includegraphics{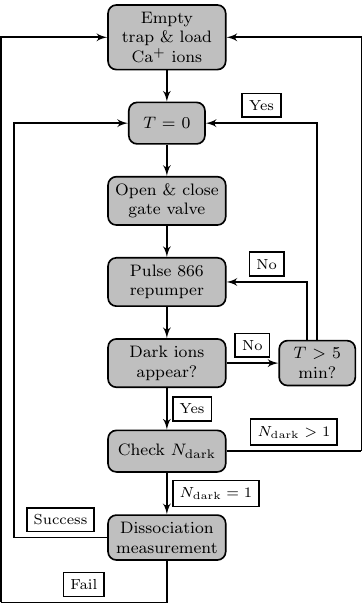}
\caption{Photodissociation measurement process tree.}
\label{fig:exp sch}
\end{figure}

The photodissociation experiments in this paper are carried out with an automated script according to the process tree illustrated in Figure.~\ref{fig:exp sch}.

\section{Ion crystal configuration identification}\label{apd:ion finder}

\begin{figure}
    \centering
    \includegraphics[width=1.0\linewidth]{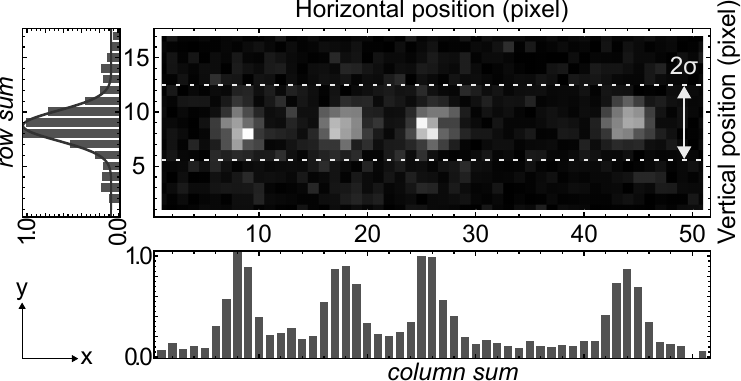}
    \caption{Shown here is a ROI in an image of 4 fluorescing Ca$^+$ and a non-fluorescing CaOH$^+$ ion. This ROI is processed into two data sets: the \textit{row sum} and \textit{column sum}, which sum and normalize the image array along the rows and columns, respectively. However, the \textit{column sum} is not accumulated over the full length of the columns but only in a $2\sigma$ wide range around the ions' vertical position, which is determined using a Gaussian fit, and $\sigma$ is the fitted waist.}
    \label{fig:line_row_sum}
\end{figure}

The photodissociation experiment requires real-time evaluation of the ion crystal configuration such that the number of fluorescing (bright) Ca$^+$ and non-fluorescing (dark) molecular ions can be monitored. Knowledge of the ion crystal configuration allows a script to run automatic measurements and take actions such as (re)loading the trap or emptying it if more than one dark ion is created. The ion crystal configuration is identified using solely image data of the ions taken by an imaging system with $10 \times$ magnification using an Andor iXon$^{\textrm{EM}}$+ 885 camera with a pixel size of 8$\,$µm. For this process, a region of interest (ROI) in the image centered around the ions is converted into two normalized data sets: the \textit{column sum}, where the ROI is summed along the columns of the image array; and the \textit{row sum}, where the summation is done along the rows (see Fig.~\ref{fig:line_row_sum}). All analysis to identify the ion crystal configuration is based on these two data sets.

The identification of the ion crystal configuration is performed in real-time and runs continuously throughout the measurements. An image of the ions is taken with an exposure time of 0.2$\,$s, from which the ion crystal configuration is identified. The identification process takes $\sim 0.2\,$s, then the next image is taken and analyzed. Three algorithms are used to determine the ion crystal configuration: (i) check for the presence of bright ions; (ii) counting of bright ions; and (iii) determination of the ion crystal configuration via a template fitting method.

The first method checks for the presence or absence of bright ions in the trap. Here, a moving average is applied to the \textit{row sum} of the image data before applying a discrete Fourier transform (DFT). If the trap is empty, the moving average is mainly flat and the amplitude of the longest spatial wavelength in the DFT is small; if there are bright ions in the trap, the moving average exhibits a broad peak, and thus the amplitude of the longest spatial wavelength in the DFT is larger. A threshold is then set to differentiate between an empty or a non-empty trap.

The number of bright ions is counted from the \textit{column sum} via a \textit{peak finder} algorithm, which can determine the number and positions of bright ions without requiring a fluorescence-based threshold. This algorithm extracts and sorts local maxima of \textit{column sum} over subsets of length comparable but smaller than the ion separation. From the largest amplitude difference and their amplitudes, the algorithm classifies peaks into two distinct groups: large-amplitude peaks originating from bright ions and small-amplitude peaks due to noise. Knowledge of the positions of the bright ion peaks can be utilized to identify the presence of dark ions for, e.g., determining the trap center position, as it requires an ion crystal of solely bright ions.

In order to count dark ions, a template fitting procedure was performed to identify the ion crystal configuration. In this process, template sets for different ion numbers and configurations are compared to the \textit{column sum} data set. Template sets are generated by the following scheme. The desired ion configuration in a template for $N$ ions is given by the vector $\mathbf{A}^N$. The value of $A_i^N$ is assigned to $1$ for a bright Ca$^+$ ion and $0$ for a dark molecular ion, for $i\in \left[1,N\right]$, e.g., the ion crystal configuration template which matches the \textit{row sum} shown in Fig.~\ref{fig:line_row_sum} is $\mathbf{A}^5=(1,1,1,0,1)$. Then the sum of Gaussian peaks
    \begin{equation}
    \label{Equ:gauss_chain}
    g(x; \mathbf{p}^N,\mathbf{A}^N,\sigma,x_0,d)=\sum_{i=1}^N A_i^N\ e^{-\frac{2}{\sigma^2}(x-p_i^N+x_0-d)^2}
    \end{equation}
is discretized by performing the set of integrals
    \begin{equation}
    T_k^N(\mathbf{A}^N)=\int_{x_k^\textrm{s}}^{x_k^\textrm{e}}g(x, \mathbf{p}^N,\mathbf{A}^N,\sigma,x_0,d)\ \mathrm{d}x.
    \end{equation}
Here $\mathbf{p}^N=(p_1,...,p_N)$ denotes the numerically calculated positions\cite{jamesQuantumDynamicsCold1998} of the ions with respect to the calibrated trap center position $x_0$, $\sigma$ is the waist of the Gaussian peaks (taken from the fit presented in Fig.~\ref{fig:line_row_sum}), and $d$ is an offset of the trap center position. The start and end positions, $x_k^\textrm{s}$ and $x_k^\textrm{e}$, of the $k$-th binning interval are chosen such that the binning and length of the template match those of the measured \textit{column sum} data set. $\mathbf{T}^N(\mathbf{A}^N)$ is the binned template of $N$ ions for the configuration $\mathbf{A}^N$.  To find the particular ion crystal configuration, a set of templates is created for all reasonable dark and bright ion numbers and all possible corresponding ion configurations. All templates are compared with the \textit{column sum} data set using the comparison function
    \begin{equation}
    C=\sum_{k=1}^L \big[T_k^N(\mathbf{A}^N)-S^\textrm{col}_k\big]^2,
    \end{equation}
where $\mathbf{S}^\textrm{col}$ denotes the \textit{column sum} data set and $L$ is its size. The best-fitting template, given by the minimum outcome of $C$, then determines the ion crystal configuration. This method reliably counts the total, bright, and dark ion numbers and thus enables automation of the photodissociation and mass spectrometry measurements. Since this method is vulnerable to drifts of the trap center position, every particular template is scanned over \textit{column sum} by varying the offset $d$. This allows tracking of the trap center position and automated recalibration when the drift exceeds a given threshold.

\section{Mass spectrometry}\label{apd:mass spec}

\begin{figure}
    \centering
    \includegraphics[width=\linewidth]{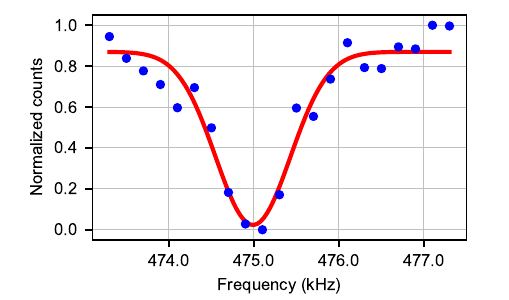}
    \caption{The normalized photon counts (blue dots) as a function of tickling frequency measured for a co-trapped Ca$^+$ and molecular ion to perform mass spectrometry. The COM motional frequency obtained from a Gaussian fit (red curve) gives a mass of 56.9(1)$\,$amu for the molecular ion which is consistent with CaOH$^+$ and excludes the nearest other possibilities, CaO$^+$ and Ca$^+$(H$_2$O).}
    \label{fig:mass_spec_fig}
\end{figure}

To determine the molecular mass, a radio frequency (RF) ``tickling'' signal with an amplitude of 30$\,$mV modulates the DC voltage on the endcap electrodes which defines the axial confinement. The signal excites the COM motion of the ions which peaks when the signal frequency is on resonance with the COM motional frequency. Due to Doppler broadening, the ions fluoresce less when their motional state is excited\cite{winelandRadiationPressureCoolingBound1978}. The COM motional frequency can then be determined by scanning the RF tickling frequency to minimize the fluorescence.

In crystals with mixed ion species, the COM motional frequency depends not only on the different ion masses but also on the specific ion crystal configuration\cite{homeQuantumScienceMetrology2013}. For instance, the two ion configurations (Ca$^+$, Ca$^+$, CaOH$^+$) and (Ca$^+$, CaOH$^+$, Ca$^+$) have a different COM motional frequency, with a deviation of 1.7$\,$kHz in our system. When applying an RF signal which is close to the COM motional frequency of the current ion configuration, the ions reconfigure their respective positions due to the induced motion. After the reconfiguration, the RF signal is no longer resonant with the COM motion which makes it difficult to perform a precise measurement of the resonance frequency. Thus, we perform measurements with crystals consisting of exactly two trapped ions, as there exists only a single COM motional frequency due to symmetry. However, the dissociation measurements are performed with larger ion crystals to speed up the CaOH$^+$ generation process. To prevent ion loss during the tickling scan, a Doppler cooling cycle of about $\sim \text{1} \, \text{s}$ is applied to the ions with the RF signal being turned off after each tickling measurement in the scan. For every single Ca$^+$ and molecular ion pair, we performed and summed up to five measurement runs.

In order to determine the COM motional frequency from the images taken with the EMCCD camera, the measured photon counts at each pixel are summed in a small ROI around the ions. This ROI is $4 \sigma$ wide along the y-axis of the images, where $\sigma$ is the waist of a Gaussian fitted to the \textit{row sum} dataset, as shown in Fig.~\ref{fig:line_row_sum}. Along the x-axis, this small ROI has a length which is 5 times that of a two ion crystal with an endcap electrode voltage of 205$\,$V, i.e., 8.5$\,$µm. The total photon counts integrated over the ROI as a function of the RF tickling frequency are normalized and the COM motional frequency is determined by fitting a Gaussian function to the data, as shown in Fig.~\ref{fig:mass_spec_fig}.

The fitted COM motional frequencies from 65 Ca$^+$ and molecular ion pairs and the corresponding reference COM motional frequencies of Ca$^+$ gave masses for the molecular ions ranging from $56.7-57.2\,$amu. This confirms that all measured 65 molecular ions were CaOH$^+$.

In order to estimate the statistical uncertainty of the measured COM motional frequency of trapped atomic Ca$^+$ ions, we evaluated 34 data sets taken over the course of hours on two different days. For a voltage of 205$\,$V applied to the endcap electrodes, a COM motional frequency of 526.8$\,$kHz with a standard deviation of 200 Hz was determined. Since the errors in the fitted COM motional frequency of all 65 processed datasets of Ca$^+$ and molecular ion pairs and all reference datasets are all smaller than 200$\,$Hz, this value is utilized to obtain an upper limit for the uncertainty of COM motional frequencies. This results in an estimated uncertainty in the derived molecular ion masses of 0.1$\,$amu.

\bibliography{Photodissociation202310_bibtex}

\begin{thebibliography}{47}%
\makeatletter
\providecommand \@ifxundefined [1]{%
 \@ifx{#1\undefined}
}%
\providecommand \@ifnum [1]{%
 \ifnum #1\expandafter \@firstoftwo
 \else \expandafter \@secondoftwo
 \fi
}%
\providecommand \@ifx [1]{%
 \ifx #1\expandafter \@firstoftwo
 \else \expandafter \@secondoftwo
 \fi
}%
\providecommand \natexlab [1]{#1}%
\providecommand \enquote  [1]{``#1''}%
\providecommand \bibnamefont  [1]{#1}%
\providecommand \bibfnamefont [1]{#1}%
\providecommand \citenamefont [1]{#1}%
\providecommand \href@noop [0]{\@secondoftwo}%
\providecommand \href [0]{\begingroup \@sanitize@url \@href}%
\providecommand \@href[1]{\@@startlink{#1}\@@href}%
\providecommand \@@href[1]{\endgroup#1\@@endlink}%
\providecommand \@sanitize@url [0]{\catcode `\\12\catcode `\$12\catcode `\&12\catcode `\#12\catcode `\^12\catcode `\_12\catcode `\%12\relax}%
\providecommand \@@startlink[1]{}%
\providecommand \@@endlink[0]{}%
\providecommand \url  [0]{\begingroup\@sanitize@url \@url }%
\providecommand \@url [1]{\endgroup\@href {#1}{\urlprefix }}%
\providecommand \urlprefix  [0]{URL }%
\providecommand \Eprint [0]{\href }%
\providecommand \doibase [0]{http://dx.doi.org/}%
\providecommand \selectlanguage [0]{\@gobble}%
\providecommand \bibinfo  [0]{\@secondoftwo}%
\providecommand \bibfield  [0]{\@secondoftwo}%
\providecommand \translation [1]{[#1]}%
\providecommand \BibitemOpen [0]{}%
\providecommand \bibitemStop [0]{}%
\providecommand \bibitemNoStop [0]{.\EOS\space}%
\providecommand \EOS [0]{\spacefactor3000\relax}%
\providecommand \BibitemShut  [1]{\csname bibitem#1\endcsname}%
\let\auto@bib@innerbib\@empty
%</preamble>
\bibitem [{\citenamefont {DeMille}(2002)}]{demilleQuantumComputationTrapped2002}%
  \BibitemOpen
  \bibfield  {author} {\bibinfo {author} {\bibfnamefont {D.}~\bibnamefont {DeMille}},\ }\bibfield  {title} {\enquote {\bibinfo {title} {Quantum {{Computation}} with {{Trapped Polar Molecules}}},}\ }\href {\doibase 10.1103/PhysRevLett.88.067901} {\bibfield  {journal} {\bibinfo  {journal} {Phys. Rev. Lett.}\ }\textbf {\bibinfo {volume} {88}},\ \bibinfo {pages} {067901} (\bibinfo {year} {2002})}\BibitemShut {NoStop}%
\bibitem [{\citenamefont {Lin}\ \emph {et~al.}(2020)\citenamefont {Lin}, \citenamefont {Leibrandt}, \citenamefont {Leibfried},\ and\ \citenamefont {Chou}}]{linQuantumEntanglementAtom2020}%
  \BibitemOpen
  \bibfield  {author} {\bibinfo {author} {\bibfnamefont {Y.}~\bibnamefont {Lin}}, \bibinfo {author} {\bibfnamefont {D.~R.}\ \bibnamefont {Leibrandt}}, \bibinfo {author} {\bibfnamefont {D.}~\bibnamefont {Leibfried}}, \ and\ \bibinfo {author} {\bibfnamefont {C.-w.}\ \bibnamefont {Chou}},\ }\bibfield  {title} {\enquote {\bibinfo {title} {Quantum entanglement between an atom and a molecule},}\ }\href {\doibase 10.1038/s41586-020-2257-1} {\bibfield  {journal} {\bibinfo  {journal} {Nature}\ }\textbf {\bibinfo {volume} {581}},\ \bibinfo {pages} {273} (\bibinfo {year} {2020})}\BibitemShut {NoStop}%
\bibitem [{\citenamefont {Roussy}\ \emph {et~al.}(2023)\citenamefont {Roussy}, \citenamefont {Caldwell}, \citenamefont {Wright}, \citenamefont {Cairncross}, \citenamefont {Shagam}, \citenamefont {Ng}, \citenamefont {Schlossberger}, \citenamefont {Park}, \citenamefont {Wang}, \citenamefont {Ye},\ and\ \citenamefont {Cornell}}]{roussyImprovedBoundElectron2023}%
  \BibitemOpen
  \bibfield  {author} {\bibinfo {author} {\bibfnamefont {T.~S.}\ \bibnamefont {Roussy}}, \bibinfo {author} {\bibfnamefont {L.}~\bibnamefont {Caldwell}}, \bibinfo {author} {\bibfnamefont {T.}~\bibnamefont {Wright}}, \bibinfo {author} {\bibfnamefont {W.~B.}\ \bibnamefont {Cairncross}}, \bibinfo {author} {\bibfnamefont {Y.}~\bibnamefont {Shagam}}, \bibinfo {author} {\bibfnamefont {K.~B.}\ \bibnamefont {Ng}}, \bibinfo {author} {\bibfnamefont {N.}~\bibnamefont {Schlossberger}}, \bibinfo {author} {\bibfnamefont {S.~Y.}\ \bibnamefont {Park}}, \bibinfo {author} {\bibfnamefont {A.}~\bibnamefont {Wang}}, \bibinfo {author} {\bibfnamefont {J.}~\bibnamefont {Ye}}, \ and\ \bibinfo {author} {\bibfnamefont {E.~A.}\ \bibnamefont {Cornell}},\ }\bibfield  {title} {\enquote {\bibinfo {title} {An improved bound on the electron's electric dipole moment},}\ }\href {\doibase 10.1126/science.adg4084} {\bibfield  {journal} {\bibinfo  {journal} {Science}\ }\textbf {\bibinfo {volume} {381}},\ \bibinfo {pages} {46} (\bibinfo {year}
  {2023})}\BibitemShut {NoStop}%
\bibitem [{\citenamefont {Alighanbari}\ \emph {et~al.}(2020)\citenamefont {Alighanbari}, \citenamefont {Giri}, \citenamefont {Constantin}, \citenamefont {Korobov},\ and\ \citenamefont {Schiller}}]{alighanbariPreciseTestQuantum2020}%
  \BibitemOpen
  \bibfield  {author} {\bibinfo {author} {\bibfnamefont {S.}~\bibnamefont {Alighanbari}}, \bibinfo {author} {\bibfnamefont {G.~S.}\ \bibnamefont {Giri}}, \bibinfo {author} {\bibfnamefont {F.~L.}\ \bibnamefont {Constantin}}, \bibinfo {author} {\bibfnamefont {V.~I.}\ \bibnamefont {Korobov}}, \ and\ \bibinfo {author} {\bibfnamefont {S.}~\bibnamefont {Schiller}},\ }\bibfield  {title} {\enquote {\bibinfo {title} {Precise test of quantum electrodynamics and determination of fundamental constants with {{HD}}+ ions},}\ }\href {\doibase 10.1038/s41586-020-2261-5} {\bibfield  {journal} {\bibinfo  {journal} {Nature}\ }\textbf {\bibinfo {volume} {581}},\ \bibinfo {pages} {152} (\bibinfo {year} {2020})}\BibitemShut {NoStop}%
\bibitem [{\citenamefont {Nguyen}\ \emph {et~al.}(2011)\citenamefont {Nguyen}, \citenamefont {Viteri}, \citenamefont {Hohenstein}, \citenamefont {Sherrill}, \citenamefont {Brown},\ and\ \citenamefont {Odom}}]{nguyenChallengesLasercoolingMolecular2011}%
  \BibitemOpen
  \bibfield  {author} {\bibinfo {author} {\bibfnamefont {J.~H.~V.}\ \bibnamefont {Nguyen}}, \bibinfo {author} {\bibfnamefont {C.~R.}\ \bibnamefont {Viteri}}, \bibinfo {author} {\bibfnamefont {E.~G.}\ \bibnamefont {Hohenstein}}, \bibinfo {author} {\bibfnamefont {C.~D.}\ \bibnamefont {Sherrill}}, \bibinfo {author} {\bibfnamefont {K.~R.}\ \bibnamefont {Brown}}, \ and\ \bibinfo {author} {\bibfnamefont {B.}~\bibnamefont {Odom}},\ }\bibfield  {title} {\enquote {\bibinfo {title} {Challenges of laser-cooling molecular ions},}\ }\href {\doibase 10.1088/1367-2630/13/6/063023} {\bibfield  {journal} {\bibinfo  {journal} {New J. Phys.}\ }\textbf {\bibinfo {volume} {13}},\ \bibinfo {pages} {063023} (\bibinfo {year} {2011})}\BibitemShut {NoStop}%
\bibitem [{\citenamefont {Ivanov}\ \emph {et~al.}(2020)\citenamefont {Ivanov}, \citenamefont {Jagau}, \citenamefont {Zhu}, \citenamefont {Hudson},\ and\ \citenamefont {Krylov}}]{ivanovSearchMolecularIons2020}%
  \BibitemOpen
  \bibfield  {author} {\bibinfo {author} {\bibfnamefont {M.~V.}\ \bibnamefont {Ivanov}}, \bibinfo {author} {\bibfnamefont {T.-C.}\ \bibnamefont {Jagau}}, \bibinfo {author} {\bibfnamefont {G.-Z.}\ \bibnamefont {Zhu}}, \bibinfo {author} {\bibfnamefont {E.~R.}\ \bibnamefont {Hudson}}, \ and\ \bibinfo {author} {\bibfnamefont {A.~I.}\ \bibnamefont {Krylov}},\ }\bibfield  {title} {\enquote {\bibinfo {title} {In search of molecular ions for optical cycling: A difficult road},}\ }\href {\doibase 10.1039/D0CP02921A} {\bibfield  {journal} {\bibinfo  {journal} {Phys. Chem. Chem. Phys.}\ }\textbf {\bibinfo {volume} {22}},\ \bibinfo {pages} {17075} (\bibinfo {year} {2020})}\BibitemShut {NoStop}%
\bibitem [{\citenamefont {W{\'o}jcik}, \citenamefont {Hudson},\ and\ \citenamefont {Krylov}(2022)}]{wojcikProspectsOpticalCycling2022}%
  \BibitemOpen
  \bibfield  {author} {\bibinfo {author} {\bibfnamefont {P.}~\bibnamefont {W{\'o}jcik}}, \bibinfo {author} {\bibfnamefont {E.~R.}\ \bibnamefont {Hudson}}, \ and\ \bibinfo {author} {\bibfnamefont {A.~I.}\ \bibnamefont {Krylov}},\ }\bibfield  {title} {\enquote {\bibinfo {title} {On the prospects of optical cycling in diatomic cations: Effects of transition metals, spin--orbit couplings, and multiple bonds},}\ }\href {\doibase 10.1080/00268976.2022.2107582} {\bibfield  {journal} {\bibinfo  {journal} {Molecular Physics}\ }\textbf {\bibinfo {volume} {121}},\ \bibinfo {pages} {e2107582} (\bibinfo {year} {2022})}\BibitemShut {NoStop}%
\bibitem [{\citenamefont {M{\o}lhave}\ and\ \citenamefont {Drewsen}(2000)}]{molhaveFormationTranslationallyCold2000}%
  \BibitemOpen
  \bibfield  {author} {\bibinfo {author} {\bibfnamefont {K.}~\bibnamefont {M{\o}lhave}}\ and\ \bibinfo {author} {\bibfnamefont {M.}~\bibnamefont {Drewsen}},\ }\bibfield  {title} {\enquote {\bibinfo {title} {Formation of translationally cold {${\mathrm{MgH}}^{+}$} and {${\mathrm{MgD}}^{+}$} molecules in an ion trap},}\ }\href {\doibase 10.1103/PhysRevA.62.011401} {\bibfield  {journal} {\bibinfo  {journal} {Phys. Rev. A}\ }\textbf {\bibinfo {volume} {62}},\ \bibinfo {pages} {011401} (\bibinfo {year} {2000})}\BibitemShut {NoStop}%
\bibitem [{\citenamefont {Rugango}\ \emph {et~al.}(2015)\citenamefont {Rugango}, \citenamefont {Goeders}, \citenamefont {Dixon}, \citenamefont {Gray}, \citenamefont {Khanyile}, \citenamefont {Shu}, \citenamefont {Clark},\ and\ \citenamefont {Brown}}]{rugangoSympatheticCoolingMolecular2015}%
  \BibitemOpen
  \bibfield  {author} {\bibinfo {author} {\bibfnamefont {R.}~\bibnamefont {Rugango}}, \bibinfo {author} {\bibfnamefont {J.~E.}\ \bibnamefont {Goeders}}, \bibinfo {author} {\bibfnamefont {T.~H.}\ \bibnamefont {Dixon}}, \bibinfo {author} {\bibfnamefont {J.~M.}\ \bibnamefont {Gray}}, \bibinfo {author} {\bibfnamefont {N.~B.}\ \bibnamefont {Khanyile}}, \bibinfo {author} {\bibfnamefont {G.}~\bibnamefont {Shu}}, \bibinfo {author} {\bibfnamefont {R.~J.}\ \bibnamefont {Clark}}, \ and\ \bibinfo {author} {\bibfnamefont {K.~R.}\ \bibnamefont {Brown}},\ }\bibfield  {title} {\enquote {\bibinfo {title} {Sympathetic cooling of molecular ion motion to the ground state},}\ }\href {\doibase 10.1088/1367-2630/17/3/035009} {\bibfield  {journal} {\bibinfo  {journal} {New J. Phys.}\ }\textbf {\bibinfo {volume} {17}},\ \bibinfo {pages} {035009} (\bibinfo {year} {2015})}\BibitemShut {NoStop}%
\bibitem [{\citenamefont {Wan}\ \emph {et~al.}(2015)\citenamefont {Wan}, \citenamefont {Gebert}, \citenamefont {Wolf},\ and\ \citenamefont {Schmidt}}]{wanEfficientSympatheticMotionalgroundstate2015}%
  \BibitemOpen
  \bibfield  {author} {\bibinfo {author} {\bibfnamefont {Y.}~\bibnamefont {Wan}}, \bibinfo {author} {\bibfnamefont {F.}~\bibnamefont {Gebert}}, \bibinfo {author} {\bibfnamefont {F.}~\bibnamefont {Wolf}}, \ and\ \bibinfo {author} {\bibfnamefont {P.~O.}\ \bibnamefont {Schmidt}},\ }\bibfield  {title} {\enquote {\bibinfo {title} {Efficient sympathetic motional-ground-state cooling of a molecular ion},}\ }\href {\doibase 10.1103/PhysRevA.91.043425} {\bibfield  {journal} {\bibinfo  {journal} {Phys. Rev. A}\ }\textbf {\bibinfo {volume} {91}},\ \bibinfo {pages} {043425} (\bibinfo {year} {2015})}\BibitemShut {NoStop}%
\bibitem [{\citenamefont {Wolf}\ \emph {et~al.}(2016)\citenamefont {Wolf}, \citenamefont {Wan}, \citenamefont {Heip}, \citenamefont {Gebert}, \citenamefont {Shi},\ and\ \citenamefont {Schmidt}}]{wolfNondestructiveStateDetection2016}%
  \BibitemOpen
  \bibfield  {author} {\bibinfo {author} {\bibfnamefont {F.}~\bibnamefont {Wolf}}, \bibinfo {author} {\bibfnamefont {Y.}~\bibnamefont {Wan}}, \bibinfo {author} {\bibfnamefont {J.~C.}\ \bibnamefont {Heip}}, \bibinfo {author} {\bibfnamefont {F.}~\bibnamefont {Gebert}}, \bibinfo {author} {\bibfnamefont {C.}~\bibnamefont {Shi}}, \ and\ \bibinfo {author} {\bibfnamefont {P.~O.}\ \bibnamefont {Schmidt}},\ }\bibfield  {title} {\enquote {\bibinfo {title} {Non-destructive state detection for quantum logic spectroscopy of molecular ions},}\ }\href {\doibase 10.1038/nature16513} {\bibfield  {journal} {\bibinfo  {journal} {Nature}\ }\textbf {\bibinfo {volume} {530}},\ \bibinfo {pages} {457} (\bibinfo {year} {2016})}\BibitemShut {NoStop}%
\bibitem [{\citenamefont {Chou}\ \emph {et~al.}(2017)\citenamefont {Chou}, \citenamefont {Kurz}, \citenamefont {Hume}, \citenamefont {Plessow}, \citenamefont {Leibrandt},\ and\ \citenamefont {Leibfried}}]{chouPreparationCoherentManipulation2017}%
  \BibitemOpen
  \bibfield  {author} {\bibinfo {author} {\bibfnamefont {C.-w.}\ \bibnamefont {Chou}}, \bibinfo {author} {\bibfnamefont {C.}~\bibnamefont {Kurz}}, \bibinfo {author} {\bibfnamefont {D.~B.}\ \bibnamefont {Hume}}, \bibinfo {author} {\bibfnamefont {P.~N.}\ \bibnamefont {Plessow}}, \bibinfo {author} {\bibfnamefont {D.~R.}\ \bibnamefont {Leibrandt}}, \ and\ \bibinfo {author} {\bibfnamefont {D.}~\bibnamefont {Leibfried}},\ }\bibfield  {title} {\enquote {\bibinfo {title} {Preparation and coherent manipulation of pure quantum states of a single molecular ion},}\ }\href {\doibase 10.1038/nature22338} {\bibfield  {journal} {\bibinfo  {journal} {Nature}\ }\textbf {\bibinfo {volume} {545}},\ \bibinfo {pages} {203} (\bibinfo {year} {2017})}\BibitemShut {NoStop}%
\bibitem [{\citenamefont {Sinhal}\ \emph {et~al.}(2020)\citenamefont {Sinhal}, \citenamefont {Meir}, \citenamefont {Najafian}, \citenamefont {Hegi},\ and\ \citenamefont {Willitsch}}]{sinhalQuantumnondemolitionStateDetection2020}%
  \BibitemOpen
  \bibfield  {author} {\bibinfo {author} {\bibfnamefont {M.}~\bibnamefont {Sinhal}}, \bibinfo {author} {\bibfnamefont {Z.}~\bibnamefont {Meir}}, \bibinfo {author} {\bibfnamefont {K.}~\bibnamefont {Najafian}}, \bibinfo {author} {\bibfnamefont {G.}~\bibnamefont {Hegi}}, \ and\ \bibinfo {author} {\bibfnamefont {S.}~\bibnamefont {Willitsch}},\ }\bibfield  {title} {\enquote {\bibinfo {title} {Quantum-nondemolition state detection and spectroscopy of single trapped molecules},}\ }\href {\doibase 10.1126/science.aaz9837} {\bibfield  {journal} {\bibinfo  {journal} {Science}\ }\textbf {\bibinfo {volume} {367}},\ \bibinfo {pages} {1213} (\bibinfo {year} {2020})}\BibitemShut {NoStop}%
\bibitem [{\citenamefont {Chou}\ \emph {et~al.}(2020)\citenamefont {Chou}, \citenamefont {Collopy}, \citenamefont {Kurz}, \citenamefont {Lin}, \citenamefont {Harding}, \citenamefont {Plessow}, \citenamefont {Fortier}, \citenamefont {Diddams}, \citenamefont {Leibfried},\ and\ \citenamefont {Leibrandt}}]{chouFrequencycombSpectroscopyPure2020}%
  \BibitemOpen
  \bibfield  {author} {\bibinfo {author} {\bibfnamefont {C.~W.}\ \bibnamefont {Chou}}, \bibinfo {author} {\bibfnamefont {A.~L.}\ \bibnamefont {Collopy}}, \bibinfo {author} {\bibfnamefont {C.}~\bibnamefont {Kurz}}, \bibinfo {author} {\bibfnamefont {Y.}~\bibnamefont {Lin}}, \bibinfo {author} {\bibfnamefont {M.~E.}\ \bibnamefont {Harding}}, \bibinfo {author} {\bibfnamefont {P.~N.}\ \bibnamefont {Plessow}}, \bibinfo {author} {\bibfnamefont {T.}~\bibnamefont {Fortier}}, \bibinfo {author} {\bibfnamefont {S.}~\bibnamefont {Diddams}}, \bibinfo {author} {\bibfnamefont {D.}~\bibnamefont {Leibfried}}, \ and\ \bibinfo {author} {\bibfnamefont {D.~R.}\ \bibnamefont {Leibrandt}},\ }\bibfield  {title} {\enquote {\bibinfo {title} {Frequency-comb spectroscopy on pure quantum states of a single molecular ion},}\ }\href {\doibase 10.1126/science.aba3628} {\bibfield  {journal} {\bibinfo  {journal} {Science}\ }\textbf {\bibinfo {volume} {367}},\ \bibinfo {pages} {1458} (\bibinfo {year} {2020})}\BibitemShut {NoStop}%
\bibitem [{\citenamefont {Koelemeij}\ \emph {et~al.}(2007)\citenamefont {Koelemeij}, \citenamefont {Roth}, \citenamefont {Wicht}, \citenamefont {Ernsting},\ and\ \citenamefont {Schiller}}]{koelemeijVibrationalSpectroscopyMathrm2007}%
  \BibitemOpen
  \bibfield  {author} {\bibinfo {author} {\bibfnamefont {J.~C.~J.}\ \bibnamefont {Koelemeij}}, \bibinfo {author} {\bibfnamefont {B.}~\bibnamefont {Roth}}, \bibinfo {author} {\bibfnamefont {A.}~\bibnamefont {Wicht}}, \bibinfo {author} {\bibfnamefont {I.}~\bibnamefont {Ernsting}}, \ and\ \bibinfo {author} {\bibfnamefont {S.}~\bibnamefont {Schiller}},\ }\bibfield  {title} {\enquote {\bibinfo {title} {Vibrational {{Spectroscopy}} of {${\mathrm{HD}}^{+}$} with 2-ppb {{Accuracy}}},}\ }\href {\doibase 10.1103/PhysRevLett.98.173002} {\bibfield  {journal} {\bibinfo  {journal} {Phys. Rev. Lett.}\ }\textbf {\bibinfo {volume} {98}},\ \bibinfo {pages} {173002} (\bibinfo {year} {2007})}\BibitemShut {NoStop}%
\bibitem [{\citenamefont {Karr}, \citenamefont {Douillet},\ and\ \citenamefont {Hilico}(2012)}]{karrPhotodissociationTrappedMathrm2012}%
  \BibitemOpen
  \bibfield  {author} {\bibinfo {author} {\bibfnamefont {J.-P.}\ \bibnamefont {Karr}}, \bibinfo {author} {\bibfnamefont {A.}~\bibnamefont {Douillet}}, \ and\ \bibinfo {author} {\bibfnamefont {L.}~\bibnamefont {Hilico}},\ }\bibfield  {title} {\enquote {\bibinfo {title} {Photodissociation of trapped {$\mathrm{H}_{2}^{+}$}ions for {{REMPD}} spectroscopy},}\ }\href {\doibase 10.1007/s00340-011-4757-z} {\bibfield  {journal} {\bibinfo  {journal} {Appl. Phys. B}\ }\textbf {\bibinfo {volume} {107}},\ \bibinfo {pages} {1043} (\bibinfo {year} {2012})}\BibitemShut {NoStop}%
\bibitem [{\citenamefont {Seck}\ \emph {et~al.}(2014)\citenamefont {Seck}, \citenamefont {Hohenstein}, \citenamefont {Lien}, \citenamefont {Stollenwerk},\ and\ \citenamefont {Odom}}]{seckRotationalStateAnalysis2014}%
  \BibitemOpen
  \bibfield  {author} {\bibinfo {author} {\bibfnamefont {C.~M.}\ \bibnamefont {Seck}}, \bibinfo {author} {\bibfnamefont {E.~G.}\ \bibnamefont {Hohenstein}}, \bibinfo {author} {\bibfnamefont {C.-Y.}\ \bibnamefont {Lien}}, \bibinfo {author} {\bibfnamefont {P.~R.}\ \bibnamefont {Stollenwerk}}, \ and\ \bibinfo {author} {\bibfnamefont {B.~C.}\ \bibnamefont {Odom}},\ }\bibfield  {title} {\enquote {\bibinfo {title} {Rotational state analysis of {{AlH}}{$^+$} by two-photon dissociation},}\ }\href {\doibase 10.1016/j.jms.2014.03.023} {\bibfield  {journal} {\bibinfo  {journal} {Journal of Molecular Spectroscopy}\ }\textbf {\bibinfo {volume} {300}},\ \bibinfo {pages} {108} (\bibinfo {year} {2014})}\BibitemShut {NoStop}%
\bibitem [{\citenamefont {Ni}\ \emph {et~al.}(2014)\citenamefont {Ni}, \citenamefont {Loh}, \citenamefont {Grau}, \citenamefont {Cossel}, \citenamefont {Ye},\ and\ \citenamefont {Cornell}}]{niStatespecificDetectionTrapped2014}%
  \BibitemOpen
  \bibfield  {author} {\bibinfo {author} {\bibfnamefont {K.-K.}\ \bibnamefont {Ni}}, \bibinfo {author} {\bibfnamefont {H.}~\bibnamefont {Loh}}, \bibinfo {author} {\bibfnamefont {M.}~\bibnamefont {Grau}}, \bibinfo {author} {\bibfnamefont {K.~C.}\ \bibnamefont {Cossel}}, \bibinfo {author} {\bibfnamefont {J.}~\bibnamefont {Ye}}, \ and\ \bibinfo {author} {\bibfnamefont {E.~A.}\ \bibnamefont {Cornell}},\ }\bibfield  {title} {\enquote {\bibinfo {title} {State-specific detection of trapped {{HfF}}+ by photodissociation},}\ }\href {\doibase https://doi.org/10.1016/j.jms.2014.02.001} {\bibfield  {journal} {\bibinfo  {journal} {Journal of Molecular Spectroscopy}\ }\textbf {\bibinfo {volume} {300}},\ \bibinfo {pages} {12} (\bibinfo {year} {2014})}\BibitemShut {NoStop}%
\bibitem [{\citenamefont {Khanyile}, \citenamefont {Shu},\ and\ \citenamefont {Brown}(2015)}]{khanyileObservationVibrationalOvertones2015}%
  \BibitemOpen
  \bibfield  {author} {\bibinfo {author} {\bibfnamefont {N.~B.}\ \bibnamefont {Khanyile}}, \bibinfo {author} {\bibfnamefont {G.}~\bibnamefont {Shu}}, \ and\ \bibinfo {author} {\bibfnamefont {K.~R.}\ \bibnamefont {Brown}},\ }\bibfield  {title} {\enquote {\bibinfo {title} {Observation of vibrational overtones by single-molecule resonant photodissociation},}\ }\href {\doibase 10.1038/ncomms8825} {\bibfield  {journal} {\bibinfo  {journal} {Nat Commun}\ }\textbf {\bibinfo {volume} {6}},\ \bibinfo {pages} {7825} (\bibinfo {year} {2015})}\BibitemShut {NoStop}%
\bibitem [{\citenamefont {Rugango}\ \emph {et~al.}(2016)\citenamefont {Rugango}, \citenamefont {Calvin}, \citenamefont {Janardan}, \citenamefont {Shu},\ and\ \citenamefont {Brown}}]{rugangoVibronicSpectroscopySympathetically2016}%
  \BibitemOpen
  \bibfield  {author} {\bibinfo {author} {\bibfnamefont {R.}~\bibnamefont {Rugango}}, \bibinfo {author} {\bibfnamefont {A.~T.}\ \bibnamefont {Calvin}}, \bibinfo {author} {\bibfnamefont {S.}~\bibnamefont {Janardan}}, \bibinfo {author} {\bibfnamefont {G.}~\bibnamefont {Shu}}, \ and\ \bibinfo {author} {\bibfnamefont {K.~R.}\ \bibnamefont {Brown}},\ }\bibfield  {title} {\enquote {\bibinfo {title} {Vibronic {{Spectroscopy}} of {{Sympathetically Cooled CaH}}+},}\ }\href {\doibase 10.1002/cphc.201600645} {\bibfield  {journal} {\bibinfo  {journal} {ChemPhysChem}\ }\textbf {\bibinfo {volume} {17}},\ \bibinfo {pages} {3764} (\bibinfo {year} {2016})}\BibitemShut {NoStop}%
\bibitem [{\citenamefont {Okada}\ \emph {et~al.}(2003)\citenamefont {Okada}, \citenamefont {Wada}, \citenamefont {Boesten}, \citenamefont {Nakamura}, \citenamefont {Katayama},\ and\ \citenamefont {Ohtani}}]{okadaAccelerationChemicalReaction2003}%
  \BibitemOpen
  \bibfield  {author} {\bibinfo {author} {\bibfnamefont {K.}~\bibnamefont {Okada}}, \bibinfo {author} {\bibfnamefont {M.}~\bibnamefont {Wada}}, \bibinfo {author} {\bibfnamefont {L.}~\bibnamefont {Boesten}}, \bibinfo {author} {\bibfnamefont {T.}~\bibnamefont {Nakamura}}, \bibinfo {author} {\bibfnamefont {I.}~\bibnamefont {Katayama}}, \ and\ \bibinfo {author} {\bibfnamefont {S.}~\bibnamefont {Ohtani}},\ }\bibfield  {title} {\enquote {\bibinfo {title} {Acceleration of the chemical reaction of trapped {{Ca}}{$^+$} ions with {{H}}{$_2$}{{O}} molecules by laser excitation},}\ }\href {\doibase 10.1088/0953-4075/36/1/304} {\bibfield  {journal} {\bibinfo  {journal} {J. Phys. B: At. Mol. Opt. Phys.}\ }\textbf {\bibinfo {volume} {36}},\ \bibinfo {pages} {33} (\bibinfo {year} {2003})}\BibitemShut {NoStop}%
\bibitem [{\citenamefont {Bertelsen}, \citenamefont {J{\o}rgensen},\ and\ \citenamefont {Drewsen}(2006)}]{bertelsenRotationalTemperaturePolar2006}%
  \BibitemOpen
  \bibfield  {author} {\bibinfo {author} {\bibfnamefont {A.}~\bibnamefont {Bertelsen}}, \bibinfo {author} {\bibfnamefont {S.}~\bibnamefont {J{\o}rgensen}}, \ and\ \bibinfo {author} {\bibfnamefont {M.}~\bibnamefont {Drewsen}},\ }\bibfield  {title} {\enquote {\bibinfo {title} {The rotational temperature of polar molecular ions in {{Coulomb}} crystals},}\ }\href {\doibase 10.1088/0953-4075/39/5/L02} {\bibfield  {journal} {\bibinfo  {journal} {J. Phys. B: At. Mol. Opt. Phys.}\ }\textbf {\bibinfo {volume} {39}},\ \bibinfo {pages} {L83} (\bibinfo {year} {2006})}\BibitemShut {NoStop}%
\bibitem [{\citenamefont {Schindler}\ \emph {et~al.}(2013)\citenamefont {Schindler}, \citenamefont {Nigg}, \citenamefont {Monz}, \citenamefont {Barreiro}, \citenamefont {Martinez}, \citenamefont {Wang}, \citenamefont {Quint}, \citenamefont {Brandl}, \citenamefont {Nebendahl}, \citenamefont {Roos}, \citenamefont {Chwalla}, \citenamefont {Hennrich},\ and\ \citenamefont {Blatt}}]{schindlerQuantumInformationProcessor2013}%
  \BibitemOpen
  \bibfield  {author} {\bibinfo {author} {\bibfnamefont {P.}~\bibnamefont {Schindler}}, \bibinfo {author} {\bibfnamefont {D.}~\bibnamefont {Nigg}}, \bibinfo {author} {\bibfnamefont {T.}~\bibnamefont {Monz}}, \bibinfo {author} {\bibfnamefont {J.~T.}\ \bibnamefont {Barreiro}}, \bibinfo {author} {\bibfnamefont {E.}~\bibnamefont {Martinez}}, \bibinfo {author} {\bibfnamefont {S.~X.}\ \bibnamefont {Wang}}, \bibinfo {author} {\bibfnamefont {S.}~\bibnamefont {Quint}}, \bibinfo {author} {\bibfnamefont {M.~F.}\ \bibnamefont {Brandl}}, \bibinfo {author} {\bibfnamefont {V.}~\bibnamefont {Nebendahl}}, \bibinfo {author} {\bibfnamefont {C.~F.}\ \bibnamefont {Roos}}, \bibinfo {author} {\bibfnamefont {M.}~\bibnamefont {Chwalla}}, \bibinfo {author} {\bibfnamefont {M.}~\bibnamefont {Hennrich}}, \ and\ \bibinfo {author} {\bibfnamefont {R.}~\bibnamefont {Blatt}},\ }\bibfield  {title} {\enquote {\bibinfo {title} {A quantum information processor with trapped ions},}\ }\href {\doibase 10.1088/1367-2630/15/12/123012} {\bibfield
  {journal} {\bibinfo  {journal} {New J. Phys.}\ }\textbf {\bibinfo {volume} {15}},\ \bibinfo {pages} {123012} (\bibinfo {year} {2013})}\BibitemShut {NoStop}%
\bibitem [{\citenamefont {Hansen}\ \emph {et~al.}(2012)\citenamefont {Hansen}, \citenamefont {S{\o}rensen}, \citenamefont {Staanum},\ and\ \citenamefont {Drewsen}}]{hansenSingleIonRecyclingReactions2012}%
  \BibitemOpen
  \bibfield  {author} {\bibinfo {author} {\bibfnamefont {A.~K.}\ \bibnamefont {Hansen}}, \bibinfo {author} {\bibfnamefont {M.~A.}\ \bibnamefont {S{\o}rensen}}, \bibinfo {author} {\bibfnamefont {P.~F.}\ \bibnamefont {Staanum}}, \ and\ \bibinfo {author} {\bibfnamefont {M.}~\bibnamefont {Drewsen}},\ }\bibfield  {title} {\enquote {\bibinfo {title} {Single-{{Ion Recycling Reactions}}},}\ }\href {\doibase 10.1002/anie.201203550} {\bibfield  {journal} {\bibinfo  {journal} {Angewandte Chemie International Edition}\ }\textbf {\bibinfo {volume} {51}},\ \bibinfo {pages} {7960} (\bibinfo {year} {2012})}\BibitemShut {NoStop}%
\bibitem [{\citenamefont {Abe}\ \emph {et~al.}(2012)\citenamefont {Abe}, \citenamefont {Moriwaki}, \citenamefont {Hada},\ and\ \citenamefont {Kajita}}]{abeInitioStudyPotential2012}%
  \BibitemOpen
  \bibfield  {author} {\bibinfo {author} {\bibfnamefont {M.}~\bibnamefont {Abe}}, \bibinfo {author} {\bibfnamefont {Y.}~\bibnamefont {Moriwaki}}, \bibinfo {author} {\bibfnamefont {M.}~\bibnamefont {Hada}}, \ and\ \bibinfo {author} {\bibfnamefont {M.}~\bibnamefont {Kajita}},\ }\bibfield  {title} {\enquote {\bibinfo {title} {Ab initio study on potential energy curves of electronic ground and excited states of {{40CaH}}+ molecule},}\ }\href {\doibase 10.1016/j.cplett.2011.11.048} {\bibfield  {journal} {\bibinfo  {journal} {Chemical Physics Letters}\ }\textbf {\bibinfo {volume} {521}},\ \bibinfo {pages} {31} (\bibinfo {year} {2012})}\BibitemShut {NoStop}%
\bibitem [{\citenamefont {VanGundy}, \citenamefont {Bartlett},\ and\ \citenamefont {Heaven}(2018)}]{vangundySpectroscopyLowlyingStates2018}%
  \BibitemOpen
  \bibfield  {author} {\bibinfo {author} {\bibfnamefont {R.~A.}\ \bibnamefont {VanGundy}}, \bibinfo {author} {\bibfnamefont {J.~H.}\ \bibnamefont {Bartlett}}, \ and\ \bibinfo {author} {\bibfnamefont {M.~C.}\ \bibnamefont {Heaven}},\ }\bibfield  {title} {\enquote {\bibinfo {title} {Spectroscopy of the low-lying states of {{CaO}}+},}\ }\href {\doibase 10.1016/j.jms.2017.10.001} {\bibfield  {journal} {\bibinfo  {journal} {Journal of Molecular Spectroscopy}\ }\textbf {\bibinfo {volume} {344}},\ \bibinfo {pages} {17} (\bibinfo {year} {2018})}\BibitemShut {NoStop}%
\bibitem [{\citenamefont {Qi}, \citenamefont {Reed},\ and\ \citenamefont {Brown}(2023)}]{qiAdiabaticallyControlledMotional2023}%
  \BibitemOpen
  \bibfield  {author} {\bibinfo {author} {\bibfnamefont {L.}~\bibnamefont {Qi}}, \bibinfo {author} {\bibfnamefont {E.~C.}\ \bibnamefont {Reed}}, \ and\ \bibinfo {author} {\bibfnamefont {K.~R.}\ \bibnamefont {Brown}},\ }\bibfield  {title} {\enquote {\bibinfo {title} {Adiabatically controlled motional states of a {${\mathrm{CaO}}^{+}$} and {${\mathrm{Ca}}^{+}$} trapped-ion chain cooled to the ground state},}\ }\href {\doibase 10.1103/PhysRevA.108.013108} {\bibfield  {journal} {\bibinfo  {journal} {Phys. Rev. A}\ }\textbf {\bibinfo {volume} {108}},\ \bibinfo {pages} {013108} (\bibinfo {year} {2023})}\BibitemShut {NoStop}%
\bibitem [{\citenamefont {Wu}\ \emph {et~al.}(2021)\citenamefont {Wu}, \citenamefont {Mills}, \citenamefont {West}, \citenamefont {Heaven},\ and\ \citenamefont {Hudson}}]{wuIncreaseBariumIontrap2021}%
  \BibitemOpen
  \bibfield  {author} {\bibinfo {author} {\bibfnamefont {H.}~\bibnamefont {Wu}}, \bibinfo {author} {\bibfnamefont {M.}~\bibnamefont {Mills}}, \bibinfo {author} {\bibfnamefont {E.}~\bibnamefont {West}}, \bibinfo {author} {\bibfnamefont {M.~C.}\ \bibnamefont {Heaven}}, \ and\ \bibinfo {author} {\bibfnamefont {E.~R.}\ \bibnamefont {Hudson}},\ }\bibfield  {title} {\enquote {\bibinfo {title} {Increase of the barium ion-trap lifetime via photodissociation},}\ }\href {\doibase 10.1103/PhysRevA.104.063103} {\bibfield  {journal} {\bibinfo  {journal} {Phys. Rev. A}\ }\textbf {\bibinfo {volume} {104}},\ \bibinfo {pages} {063103} (\bibinfo {year} {2021})}\BibitemShut {NoStop}%
\bibitem [{\citenamefont {Okada}\ \emph {et~al.}(2006)\citenamefont {Okada}, \citenamefont {Wada}, \citenamefont {Nakamura}, \citenamefont {Takayanagi}, \citenamefont {Katayama},\ and\ \citenamefont {Ohtani}}]{okadaPhotodissociationCaOHRegeneration2006}%
  \BibitemOpen
  \bibfield  {author} {\bibinfo {author} {\bibfnamefont {K.}~\bibnamefont {Okada}}, \bibinfo {author} {\bibfnamefont {M.}~\bibnamefont {Wada}}, \bibinfo {author} {\bibfnamefont {T.}~\bibnamefont {Nakamura}}, \bibinfo {author} {\bibfnamefont {T.}~\bibnamefont {Takayanagi}}, \bibinfo {author} {\bibfnamefont {I.}~\bibnamefont {Katayama}}, \ and\ \bibinfo {author} {\bibfnamefont {S.}~\bibnamefont {Ohtani}},\ }\bibfield  {title} {\enquote {\bibinfo {title} {Photodissociation of {{CaOH}}{$^+$} for {{Regeneration}} of {{Ca}}{$^+$} in {{Linear Paul Trap}}},}\ }\href {\doibase 10.1143/JJAP.45.956} {\bibfield  {journal} {\bibinfo  {journal} {Jpn. J. Appl. Phys.}\ }\textbf {\bibinfo {volume} {45}},\ \bibinfo {pages} {956} (\bibinfo {year} {2006})}\BibitemShut {NoStop}%
\bibitem [{\citenamefont {Pogorelov}\ \emph {et~al.}(2021)\citenamefont {Pogorelov}, \citenamefont {Feldker}, \citenamefont {Marciniak}, \citenamefont {Postler}, \citenamefont {Jacob}, \citenamefont {Krieglsteiner}, \citenamefont {Podlesnic}, \citenamefont {Meth}, \citenamefont {Negnevitsky}, \citenamefont {Stadler}, \citenamefont {H{\"o}fer}, \citenamefont {W{\"a}chter}, \citenamefont {Lakhmanskiy}, \citenamefont {Blatt}, \citenamefont {Schindler},\ and\ \citenamefont {Monz}}]{pogorelovCompactIonTrapQuantum2021}%
  \BibitemOpen
  \bibfield  {author} {\bibinfo {author} {\bibfnamefont {I.}~\bibnamefont {Pogorelov}}, \bibinfo {author} {\bibfnamefont {T.}~\bibnamefont {Feldker}}, \bibinfo {author} {\bibfnamefont {{\relax Ch}.~D.}\ \bibnamefont {Marciniak}}, \bibinfo {author} {\bibfnamefont {L.}~\bibnamefont {Postler}}, \bibinfo {author} {\bibfnamefont {G.}~\bibnamefont {Jacob}}, \bibinfo {author} {\bibfnamefont {O.}~\bibnamefont {Krieglsteiner}}, \bibinfo {author} {\bibfnamefont {V.}~\bibnamefont {Podlesnic}}, \bibinfo {author} {\bibfnamefont {M.}~\bibnamefont {Meth}}, \bibinfo {author} {\bibfnamefont {V.}~\bibnamefont {Negnevitsky}}, \bibinfo {author} {\bibfnamefont {M.}~\bibnamefont {Stadler}}, \bibinfo {author} {\bibfnamefont {B.}~\bibnamefont {H{\"o}fer}}, \bibinfo {author} {\bibfnamefont {C.}~\bibnamefont {W{\"a}chter}}, \bibinfo {author} {\bibfnamefont {K.}~\bibnamefont {Lakhmanskiy}}, \bibinfo {author} {\bibfnamefont {R.}~\bibnamefont {Blatt}}, \bibinfo {author} {\bibfnamefont {P.}~\bibnamefont {Schindler}}, \ and\ \bibinfo
  {author} {\bibfnamefont {T.}~\bibnamefont {Monz}},\ }\bibfield  {title} {\enquote {\bibinfo {title} {Compact {{Ion-Trap Quantum Computing Demonstrator}}},}\ }\href {\doibase 10.1103/PRXQuantum.2.020343} {\bibfield  {journal} {\bibinfo  {journal} {PRX Quantum}\ }\textbf {\bibinfo {volume} {2}},\ \bibinfo {pages} {020343} (\bibinfo {year} {2021})}\BibitemShut {NoStop}%
\bibitem [{\citenamefont {Drewsen}\ \emph {et~al.}(2004)\citenamefont {Drewsen}, \citenamefont {Mortensen}, \citenamefont {Martinussen}, \citenamefont {Staanum},\ and\ \citenamefont {S{\o}rensen}}]{drewsenNondestructiveIdentificationCold2004}%
  \BibitemOpen
  \bibfield  {author} {\bibinfo {author} {\bibfnamefont {M.}~\bibnamefont {Drewsen}}, \bibinfo {author} {\bibfnamefont {A.}~\bibnamefont {Mortensen}}, \bibinfo {author} {\bibfnamefont {R.}~\bibnamefont {Martinussen}}, \bibinfo {author} {\bibfnamefont {P.}~\bibnamefont {Staanum}}, \ and\ \bibinfo {author} {\bibfnamefont {J.~L.}\ \bibnamefont {S{\o}rensen}},\ }\bibfield  {title} {\enquote {\bibinfo {title} {Nondestructive {{Identification}} of {{Cold}} and {{Extremely Localized Single Molecular Ions}}},}\ }\href {\doibase 10.1103/PhysRevLett.93.243201} {\bibfield  {journal} {\bibinfo  {journal} {Phys. Rev. Lett.}\ }\textbf {\bibinfo {volume} {93}},\ \bibinfo {pages} {243201} (\bibinfo {year} {2004})}\BibitemShut {NoStop}%
\bibitem [{\citenamefont {Kielpinski}\ \emph {et~al.}(2000)\citenamefont {Kielpinski}, \citenamefont {King}, \citenamefont {Myatt}, \citenamefont {Sackett}, \citenamefont {Turchette}, \citenamefont {Itano}, \citenamefont {Monroe}, \citenamefont {Wineland},\ and\ \citenamefont {Zurek}}]{kielpinskiSympatheticCoolingTrapped2000}%
  \BibitemOpen
  \bibfield  {author} {\bibinfo {author} {\bibfnamefont {D.}~\bibnamefont {Kielpinski}}, \bibinfo {author} {\bibfnamefont {B.~E.}\ \bibnamefont {King}}, \bibinfo {author} {\bibfnamefont {C.~J.}\ \bibnamefont {Myatt}}, \bibinfo {author} {\bibfnamefont {C.~A.}\ \bibnamefont {Sackett}}, \bibinfo {author} {\bibfnamefont {Q.~A.}\ \bibnamefont {Turchette}}, \bibinfo {author} {\bibfnamefont {W.~M.}\ \bibnamefont {Itano}}, \bibinfo {author} {\bibfnamefont {C.}~\bibnamefont {Monroe}}, \bibinfo {author} {\bibfnamefont {D.~J.}\ \bibnamefont {Wineland}}, \ and\ \bibinfo {author} {\bibfnamefont {W.~H.}\ \bibnamefont {Zurek}},\ }\bibfield  {title} {\enquote {\bibinfo {title} {Sympathetic cooling of trapped ions for quantum logic},}\ }\href {\doibase 10.1103/PhysRevA.61.032310} {\bibfield  {journal} {\bibinfo  {journal} {Phys. Rev. A}\ }\textbf {\bibinfo {volume} {61}},\ \bibinfo {pages} {032310} (\bibinfo {year} {2000})}\BibitemShut {NoStop}%
\bibitem [{\citenamefont {Hampel}, \citenamefont {Peterson},\ and\ \citenamefont {Werner}(1992)}]{HampelCPL92}%
  \BibitemOpen
  \bibfield  {author} {\bibinfo {author} {\bibfnamefont {C.}~\bibnamefont {Hampel}}, \bibinfo {author} {\bibfnamefont {K.~A.}\ \bibnamefont {Peterson}}, \ and\ \bibinfo {author} {\bibfnamefont {H.-J.}\ \bibnamefont {Werner}},\ }\bibfield  {title} {\enquote {\bibinfo {title} {{A comparison of the efficiency and accuracy of the quadratic configuration interaction (QCISD), coupled cluster (CCSD), and Brueckner coupled cluster (BCCD) methods}},}\ }\href {\doibase 10.1016/0009-2614(92)86093-W} {\bibfield  {journal} {\bibinfo  {journal} {Chem. Phys. Lett.}\ }\textbf {\bibinfo {volume} {190}},\ \bibinfo {pages} {1} (\bibinfo {year} {1992})}\BibitemShut {NoStop}%
\bibitem [{\citenamefont {Werner}\ and\ \citenamefont {Knowles}(1988)}]{WernerJCP88}%
  \BibitemOpen
  \bibfield  {author} {\bibinfo {author} {\bibfnamefont {H.}~\bibnamefont {Werner}}\ and\ \bibinfo {author} {\bibfnamefont {P.~J.}\ \bibnamefont {Knowles}},\ }\bibfield  {title} {\enquote {\bibinfo {title} {{An efficient internally contracted multiconfiguration–reference configuration interaction method}},}\ }\href {\doibase 10.1063/1.455556} {\bibfield  {journal} {\bibinfo  {journal} {J. Chem. Phys.}\ }\textbf {\bibinfo {volume} {89}},\ \bibinfo {pages} {5803} (\bibinfo {year} {1988})}\BibitemShut {NoStop}%
\bibitem [{\citenamefont {Hill}\ and\ \citenamefont {Peterson}(2017)}]{HillJCP17}%
  \BibitemOpen
  \bibfield  {author} {\bibinfo {author} {\bibfnamefont {J.~G.}\ \bibnamefont {Hill}}\ and\ \bibinfo {author} {\bibfnamefont {K.~A.}\ \bibnamefont {Peterson}},\ }\bibfield  {title} {\enquote {\bibinfo {title} {{Gaussian basis sets for use in correlated molecular calculations. XI. Pseudopotential-based and all-electron relativistic basis sets for alkali metal (K--Fr) and alkaline earth (Ca--Ra) elements}},}\ }\href {\doibase 10.1063/1.5010587} {\bibfield  {journal} {\bibinfo  {journal} {J. Chem. Phys.}\ }\textbf {\bibinfo {volume} {147}},\ \bibinfo {pages} {244106} (\bibinfo {year} {2017})}\BibitemShut {NoStop}%
\bibitem [{\citenamefont {Peterson}\ and\ \citenamefont {Dunning}(2002)}]{PetersonJCP02}%
  \BibitemOpen
  \bibfield  {author} {\bibinfo {author} {\bibfnamefont {K.~A.}\ \bibnamefont {Peterson}}\ and\ \bibinfo {author} {\bibfnamefont {J.}~\bibnamefont {Dunning}, \bibfnamefont {Thom~H.}},\ }\bibfield  {title} {\enquote {\bibinfo {title} {{Accurate correlation consistent basis sets for molecular core–valence correlation effects: The second row atoms Al–Ar, and the first row atoms B–Ne revisited}},}\ }\href {\doibase 10.1063/1.1520138} {\bibfield  {journal} {\bibinfo  {journal} {J. Chem. Phys.}\ }\textbf {\bibinfo {volume} {117}},\ \bibinfo {pages} {10548} (\bibinfo {year} {2002})}\BibitemShut {NoStop}%
\bibitem [{\citenamefont {Dunning}(1989)}]{DunningJCP89}%
  \BibitemOpen
  \bibfield  {author} {\bibinfo {author} {\bibfnamefont {J.}~\bibnamefont {Dunning}, \bibfnamefont {Thom~H.}},\ }\bibfield  {title} {\enquote {\bibinfo {title} {{Gaussian basis sets for use in correlated molecular calculations. I. The atoms boron through neon and hydrogen}},}\ }\href {\doibase 10.1063/1.456153} {\bibfield  {journal} {\bibinfo  {journal} {J. Chem. Phys.}\ }\textbf {\bibinfo {volume} {90}},\ \bibinfo {pages} {1007} (\bibinfo {year} {1989})}\BibitemShut {NoStop}%
\bibitem [{\citenamefont {Lim}, \citenamefont {Stoll},\ and\ \citenamefont {Schwerdtfeger}(2006)}]{LimJCP06}%
  \BibitemOpen
  \bibfield  {author} {\bibinfo {author} {\bibfnamefont {I.~S.}\ \bibnamefont {Lim}}, \bibinfo {author} {\bibfnamefont {H.}~\bibnamefont {Stoll}}, \ and\ \bibinfo {author} {\bibfnamefont {P.}~\bibnamefont {Schwerdtfeger}},\ }\bibfield  {title} {\enquote {\bibinfo {title} {Relativistic small-core energy-consistent pseudopotentials for the alkaline-earth elements from {Ca to Ra}},}\ }\href {\doibase 10.1063/1.2148945} {\bibfield  {journal} {\bibinfo  {journal} {J. Chem. Phys.}\ }\textbf {\bibinfo {volume} {124}},\ \bibinfo {pages} {034107} (\bibinfo {year} {2006})}\BibitemShut {NoStop}%
\bibitem [{\citenamefont {Werner}\ \emph {et~al.}(2012{\natexlab{a}})\citenamefont {Werner}, \citenamefont {Knowles}, \citenamefont {R.~Lindh}, \citenamefont {Sch{\"u}tz}, \citenamefont {Celani}, \citenamefont {Korona}, \citenamefont {Rauhut}, \citenamefont {Amos}, \citenamefont {Bernhardsson}, \citenamefont {Berning}, \citenamefont {Cooper}, \citenamefont {Deegan}, \citenamefont {Dobbyn}, \citenamefont {F.~Eckert}, \citenamefont {Hampel}, \citenamefont {Hetzer}, \citenamefont {Lloyd}, \citenamefont {McNicholas}, \citenamefont {Meyer}, \citenamefont {Mura}, \citenamefont {Nicklass}, \citenamefont {Palmieri}, \citenamefont {Pitzer}, \citenamefont {Schumann}, \citenamefont {Stoll}, \citenamefont {Stone}, \citenamefont {Tarroni}, \citenamefont {Thorsteinsson}, \citenamefont {Wang},\ and\ \citenamefont {Wolf}}]{Molpro}%
  \BibitemOpen
  \bibfield  {author} {\bibinfo {author} {\bibfnamefont {H.-J.}\ \bibnamefont {Werner}}, \bibinfo {author} {\bibfnamefont {P.~J.}\ \bibnamefont {Knowles}}, \bibinfo {author} {\bibfnamefont {F.~R.~M.}\ \bibnamefont {R.~Lindh}}, \bibinfo {author} {\bibfnamefont {M.}~\bibnamefont {Sch{\"u}tz}}, \bibinfo {author} {\bibfnamefont {P.}~\bibnamefont {Celani}}, \bibinfo {author} {\bibfnamefont {T.}~\bibnamefont {Korona}}, \bibinfo {author} {\bibfnamefont {G.}~\bibnamefont {Rauhut}}, \bibinfo {author} {\bibfnamefont {R.~D.}\ \bibnamefont {Amos}}, \bibinfo {author} {\bibfnamefont {A.}~\bibnamefont {Bernhardsson}}, \bibinfo {author} {\bibfnamefont {A.}~\bibnamefont {Berning}}, \bibinfo {author} {\bibfnamefont {D.~L.}\ \bibnamefont {Cooper}}, \bibinfo {author} {\bibfnamefont {M.~J.~O.}\ \bibnamefont {Deegan}}, \bibinfo {author} {\bibfnamefont {A.~J.}\ \bibnamefont {Dobbyn}}, \bibinfo {author} {\bibfnamefont {E.~G.}\ \bibnamefont {F.~Eckert}}, \bibinfo {author} {\bibfnamefont {C.}~\bibnamefont {Hampel}}, \bibinfo {author}
  {\bibfnamefont {G.}~\bibnamefont {Hetzer}}, \bibinfo {author} {\bibfnamefont {A.~W.}\ \bibnamefont {Lloyd}}, \bibinfo {author} {\bibfnamefont {S.~J.}\ \bibnamefont {McNicholas}}, \bibinfo {author} {\bibfnamefont {W.}~\bibnamefont {Meyer}}, \bibinfo {author} {\bibfnamefont {M.~E.}\ \bibnamefont {Mura}}, \bibinfo {author} {\bibfnamefont {A.}~\bibnamefont {Nicklass}}, \bibinfo {author} {\bibfnamefont {P.}~\bibnamefont {Palmieri}}, \bibinfo {author} {\bibfnamefont {R.}~\bibnamefont {Pitzer}}, \bibinfo {author} {\bibfnamefont {U.}~\bibnamefont {Schumann}}, \bibinfo {author} {\bibfnamefont {H.}~\bibnamefont {Stoll}}, \bibinfo {author} {\bibfnamefont {A.~J.}\ \bibnamefont {Stone}}, \bibinfo {author} {\bibfnamefont {R.}~\bibnamefont {Tarroni}}, \bibinfo {author} {\bibfnamefont {T.}~\bibnamefont {Thorsteinsson}}, \bibinfo {author} {\bibfnamefont {M.}~\bibnamefont {Wang}}, \ and\ \bibinfo {author} {\bibfnamefont {A.}~\bibnamefont {Wolf}},\ }\href@noop {} {\emph {\bibinfo {title} {MOLPRO, version 2012.1, a package of
  ab initio programs}}} (\bibinfo {year} {2012}{\natexlab{a}}),\ \bibinfo {note} {see http://www.molpro.net}\BibitemShut {NoStop}%
\bibitem [{\citenamefont {Werner}\ \emph {et~al.}(2012{\natexlab{b}})\citenamefont {Werner}, \citenamefont {Knowles}, \citenamefont {Knizia}, \citenamefont {Manby},\ and\ \citenamefont {Sch{\"u}tz}}]{MOLPRO-WIREs}%
  \BibitemOpen
  \bibfield  {author} {\bibinfo {author} {\bibfnamefont {H.-J.}\ \bibnamefont {Werner}}, \bibinfo {author} {\bibfnamefont {P.~J.}\ \bibnamefont {Knowles}}, \bibinfo {author} {\bibfnamefont {G.}~\bibnamefont {Knizia}}, \bibinfo {author} {\bibfnamefont {F.~R.}\ \bibnamefont {Manby}}, \ and\ \bibinfo {author} {\bibfnamefont {M.}~\bibnamefont {Sch{\"u}tz}},\ }\bibfield  {title} {\enquote {\bibinfo {title} {{Molpro: a general-purpose quantum chemistry program package}},}\ }\href {\doibase 10.1002/wcms.82} {\bibfield  {journal} {\bibinfo  {journal} {WIREs Comput. Mol. Sci.}\ }\textbf {\bibinfo {volume} {2}},\ \bibinfo {pages} {242} (\bibinfo {year} {2012}{\natexlab{b}})}\BibitemShut {NoStop}%
\bibitem [{\citenamefont {Gronowski}, \citenamefont {Koza},\ and\ \citenamefont {Tomza}(2020)}]{GronowskiPRA20}%
  \BibitemOpen
  \bibfield  {author} {\bibinfo {author} {\bibfnamefont {M.}~\bibnamefont {Gronowski}}, \bibinfo {author} {\bibfnamefont {A.~M.}\ \bibnamefont {Koza}}, \ and\ \bibinfo {author} {\bibfnamefont {M.}~\bibnamefont {Tomza}},\ }\bibfield  {title} {\enquote {\bibinfo {title} {{Ab initio properties of the NaLi molecule in the $a^3\Sigma^+$ electronic state}},}\ }\href {\doibase 10.1103/PhysRevA.102.020801} {\bibfield  {journal} {\bibinfo  {journal} {Phys. Rev. A}\ }\textbf {\bibinfo {volume} {102}},\ \bibinfo {pages} {020801} (\bibinfo {year} {2020})}\BibitemShut {NoStop}%
\bibitem [{\citenamefont {Huber}\ and\ \citenamefont {Herzberg}(1979)}]{Huber1979}%
  \BibitemOpen
  \bibfield  {author} {\bibinfo {author} {\bibfnamefont {K.~P.}\ \bibnamefont {Huber}}\ and\ \bibinfo {author} {\bibfnamefont {G.}~\bibnamefont {Herzberg}},\ }\href {\doibase 10.1007/978-1-4757-0961-2} {\emph {\bibinfo {title} {Molecular spectra and molecular structure: IV. Constants of diatomic molecules}}}\ (\bibinfo  {publisher} {Springer},\ \bibinfo {year} {1979})\BibitemShut {NoStop}%
\bibitem [{\citenamefont {Drouin}(2013)}]{DrouinJPCA13}%
  \BibitemOpen
  \bibfield  {author} {\bibinfo {author} {\bibfnamefont {B.~J.}\ \bibnamefont {Drouin}},\ }\bibfield  {title} {\enquote {\bibinfo {title} {Isotopic spectra of the hydroxyl radical},}\ }\href {\doibase 10.1021/jp400923z} {\bibfield  {journal} {\bibinfo  {journal} {J. Phys. Chem. A}\ }\textbf {\bibinfo {volume} {117}},\ \bibinfo {pages} {10076} (\bibinfo {year} {2013})}\BibitemShut {NoStop}%
\bibitem [{\citenamefont {Schindler}(2019)}]{schindlerUltrafastInfraredSpectroscopy2019}%
  \BibitemOpen
  \bibfield  {author} {\bibinfo {author} {\bibfnamefont {P.}~\bibnamefont {Schindler}},\ }\bibfield  {title} {\enquote {\bibinfo {title} {Ultrafast infrared spectroscopy with single molecular ions},}\ }\href {\doibase 10.1088/1367-2630/ab3549} {\bibfield  {journal} {\bibinfo  {journal} {New J. Phys.}\ }\textbf {\bibinfo {volume} {21}},\ \bibinfo {pages} {083025} (\bibinfo {year} {2019})}\BibitemShut {NoStop}%
\bibitem [{\citenamefont {James}(1998)}]{jamesQuantumDynamicsCold1998}%
  \BibitemOpen
  \bibfield  {author} {\bibinfo {author} {\bibfnamefont {D.}~\bibnamefont {James}},\ }\bibfield  {title} {\enquote {\bibinfo {title} {Quantum dynamics of cold trapped ions with application to quantum computation},}\ }\href {\doibase 10.1007/s003400050373} {\bibfield  {journal} {\bibinfo  {journal} {Applied Physics B: Lasers and Optics}\ }\textbf {\bibinfo {volume} {66}},\ \bibinfo {pages} {181} (\bibinfo {year} {1998})}\BibitemShut {NoStop}%
\bibitem [{\citenamefont {Wineland}, \citenamefont {Drullinger},\ and\ \citenamefont {Walls}(1978)}]{winelandRadiationPressureCoolingBound1978}%
  \BibitemOpen
  \bibfield  {author} {\bibinfo {author} {\bibfnamefont {D.~J.}\ \bibnamefont {Wineland}}, \bibinfo {author} {\bibfnamefont {R.~E.}\ \bibnamefont {Drullinger}}, \ and\ \bibinfo {author} {\bibfnamefont {F.~L.}\ \bibnamefont {Walls}},\ }\bibfield  {title} {\enquote {\bibinfo {title} {Radiation-{{Pressure Cooling}} of {{Bound Resonant Absorbers}}},}\ }\href {\doibase 10.1103/PhysRevLett.40.1639} {\bibfield  {journal} {\bibinfo  {journal} {Phys. Rev. Lett.}\ }\textbf {\bibinfo {volume} {40}},\ \bibinfo {pages} {1639} (\bibinfo {year} {1978})}\BibitemShut {NoStop}%
\bibitem [{\citenamefont {Home}(2013)}]{homeQuantumScienceMetrology2013}%
  \BibitemOpen
  \bibfield  {author} {\bibinfo {author} {\bibfnamefont {J.~P.}\ \bibnamefont {Home}},\ }\bibfield  {title} {\enquote {\bibinfo {title} {Quantum science and metrology with mixed-species ion chains},}\ }\href {\doibase https://doi.org/10.1016/B978-0-12-408090-4.00004-9} {\bibfield  {journal} {\bibinfo  {journal} {Advances in Atomic, Molecular, and Optical Physics}\ }\textbf {\bibinfo {volume} {62}},\ \bibinfo {pages} {231} (\bibinfo {year} {2013})}\BibitemShut {NoStop}%
\end{thebibliography}%

\end{document}